\documentclass[10pt,prd,aps,nofootinbib
]{revtex4}
\usepackage{epsfig}
\usepackage{amsmath}
\usepackage{calrsfs}
\usepackage{nicefrac}
\usepackage{esint}

\def\barr{\left(\begin{array}{c}}
\def\earr{\end{array}\right)}
\def\bmat{\left(\begin{array}{cc}}
\def\emat{\end{array}\right)}

\def\a {\alpha}

\def\b {\beta}

\def\l {\lambda} 
\def\m {\mu}
\def\n {\nu}

\def\r {\rho}

\def\si{\sigma}

\def\beq{\begin{equation}}
\def\eeq{\end{equation}}
\def\bea{\begin{eqnarray}}
\def\eea{\end{eqnarray}}
\def\beqa{\begin{equation}\begin{array}{l}}
\def\eeqa{\end{array}\end{equation}}

\def\pd{\partial}

\def\nn{\nonumber}

\begin{document}

\title{Light-by-light scattering sum rules constraining meson transition
form factors}

 \author{Vladimir Pascalutsa}

\author{Vladyslav Pauk}

\author{Marc Vanderhaeghen}

\affiliation{Institut f\"ur Kernphysik, Johannes Gutenberg Universit\"at, Mainz D-55099, Germany}

\date{\today}

\begin{abstract}

Relating the forward light-by-light scattering to energy weighted integrals of
the $\gamma^\ast \gamma$ fusion cross sections, 
with one real photon ($\gamma$) and one virtual photon ($\gamma^\ast$),
we find two new exact super-convergence relations. They complement
the known super-convergence relation based on the extension of the GDH
sum rule to the light-light system. We also find a set of sum rules for the low-energy
photon-photon interaction. All of the new relations are verified here exactly at
leading order in scalar and spinor QED. The super-convergence relations,
applied to the $\gamma^\ast \gamma$ production of mesons,
lead to intricate relations between the $\gamma \gamma$ decay widths or the $\gamma^\ast \gamma$ transition form factors 
for (pseudo-) scalar, axial-vector and tensor mesons. 
We discuss the phenomenological implications of these results for mesons in both 
the light-quark sector and the charm-quark sector.

\end{abstract}

\maketitle

\newpage
\tableofcontents

\section{Introduction}

Light-by-light (LbL) scattering is a prediction of the quantum
theory \cite{Heisenberg:1935qt,Karplus:1950zza} which thus-far has
not been directly observed, mainly due to smallness of the cross section. 
On the other hand, the process of $\gamma^\ast \gamma^\ast$ fusion (by quasi-real photons $\gamma$ 
or virtual photons $\gamma^\ast$) 
into leptons and hadrons
has been observed at nearly all high-energy colliders, see e.g.~\cite{Budnev:1974de,Poppe:1986dq,Brodsky:2005wk} for reviews. The two phenomena --- LbL scattering and $\gamma\gamma$ fusion --- must be related by causality, similar to how the refraction index of light is related to its absorption in the Kramers-Kronig relation. The main goal of this work is to establish such relations and use them 
to investigate the structure of hadrons in the realm of quantum chromo-dynamics (QCD).

The electromagnetic interaction provides a clean probe and the two-photon state allows to produce hadrons with 
nearly all quantum numbers (with $C = +$), in contrast to 
the well studied single-photon scattering or production processes, 
which only accesses the vector states.  
When producing exclusive final states such as in the $\gamma^\ast \gamma^\ast \to {\rm meson}$ process, 
one accesses meson transition form factors (FFs), which are some of the 
simplest observables where the approach to the asymptotic limit of QCD is studied
along with the quark content of mesons described by distribution amplitudes (DAs).  The non-perturbative dynamics of QCD
is also playing a profound role in these FFs at low momentum transfers. 
For example, 
the transition FFs of the $\eta$ and $\eta^\prime$ mesons depend on the interplay of various symmetry breaking mechanisms in QCD, i.e.:
$U_A(1)$ symmetry breaking~\cite{Shore:2007yn}, dynamical and explicit chiral symmetry breaking. 
In addition, the $\gamma^\ast \gamma^\ast \to {\rm meson}$ transition FFs
are important for providing and improving constraints on the light-by-light hadronic 
contribution to the anomalous magnetic moment of the muon, $(g - 2)_\mu$. The hadronic contributions to 
$(g - 2)_\mu$ are at present 
the major uncertainty in the search for new, beyond Standard Model, physics in this high-precision 
quantity~\cite{Jegerlehner:2009ry}.

In recent years, new experiments at high luminosity $e^+ e^-$ colliders such as BABAR and Belle have 
vastly expanded the field of $\gamma \gamma$ physics. The result of a measurement of 
the $\gamma^\ast \gamma \to \pi^0$ FF at large momentum transfers by the 
BABAR Collaboration~\cite{Aubert:2009mc} came as a surprise, as this form factor seems to rise much 
faster than the perturbative QCD predictions for momentum transfers up to 40 GeV$^2$.  
A $\gamma \gamma$ physics program is  planned now by the BES-III Collaboration~\cite{Asner:2008nq}, which 
will allow to provide high-statistics results at intermediate momentum transfers for 
a multitude of $\gamma^\ast \gamma^\ast \to {\rm hadron}$ observables. 

In this work we use the dispersion theory to relate the two phenomena of LbL scattering and $\gamma^\ast \gamma$ 
fusion, and express the low-energy LbL scattering as integrals over the $\gamma^\ast \gamma$-fusion
cross sections, where one photon is real while the second may have arbitrary (space-like) virtuality. 
These integrals, or `sum rules', lead to 
interesting constraints on $\gamma \gamma$ decay widths or $\gamma^\ast \gamma$ transition FFs 
of $q\bar q$ states, and more general meson states. 
The first sum rule of this type involves the helicity-difference
cross-section for real photons and reads as:
\begin{eqnarray}
 \int  \limits_{s_0}^\infty \frac{ds}{s} \, \Big[ \sigma_2(s) - \sigma_0(s) \Big] =0,
 \label{eq:SR0}
\end{eqnarray}
where $s$ is the total energy squared, $s_0$ is the first inelastic threshold for the $\gamma \gamma$ fusion process, 
and the subscripts $0$ or $2$ for the $\gamma \gamma$ cross sections 
indicate the total helicity of the state of two circularly polarized photons. 
This sum rule was originally\footnote{An earlier version of this sum rule had been
proposed in Ref.~\cite{Roy:1974fz}, where a contribution from 
$\pi^0$ production appears on the right-hand side ({\em rhs}) of Eq.~(\ref{eq:SR0}), while integration on the {\em lhs} starts at the 2$\pi$ production threshold. That version would be fully compatible  with Eq.~(\ref{eq:SR0}), if it were
not for the sign of the $\pi^0$ contribution obtained in \cite{Roy:1974fz}.} inferred~\cite{Gerasimov:1973ja,Brodsky:1995fj} from the
the Gerasimov--Drell--Hearn (GDH) sum rule, using the fact that the photon
has no anomalous moments.

Parameterizing the lowest energy LbL interaction by means of an effective Lagrangian 
(which contains operators of dimension eight at lowest order) as 
\begin{equation}
\mathcal{L}^{(8)} = c_1 (F_{\mu\nu}F^{\mu\nu})^2 + c_2 (F_{\mu\nu}\tilde F^{\mu\nu})^2, 
\end{equation}
with $F$ and $\tilde F$ being the electromagnetic field strength and its dual, one finds
sum rules for the LbL low-energy constants (LECs)~\cite{Pascalutsa:2010sj}:
\begin{eqnarray}
c_1 = \frac{1}{8 \pi }\int\limits_{s_0}^{\infty} {\rm d} s\,  \frac{ \sigma_\parallel (s)}{s^2}\, , 
\quad \quad \quad c_2 = \frac{1}{8 \pi }\int\limits_{s_0}^{\infty} {\rm d} s\,  \frac{\sigma_\perp(s)}{s^2} \, ,
\end{eqnarray}
where the subscripts $||$ or $\perp$ indicate if the colliding photons
are polarized parallel or perpendicular to each other. While the GDH-type
sum rule provides a stringent constraint on the polarized
$\gamma \gamma$ fusion, the sum rules for the LECs allow one in principle to fully determine the low-energy LbL interaction 
through measuring the linearly polarized $\gamma \gamma$ fusion.

In this work we extend the GDH type sum rule to the case where one of the colliding photons is virtual, with arbitrary (space-like) virtuality. 
Furthermore, we find two additional sum rules, involving
the longitudinally polarized $\gamma^\ast \gamma$ cross sections. 
All details of sum rule derivation are gathered  in Sec.~\ref{sec2}.
In Sec.~\ref{pert}, all of the newly derived sum rules are
verified at leading order in scalar and spinor quantum electrodynamcis (QED).
Next we apply these results to the $\gamma^\ast \gamma^\ast$ fusion to mesons. 
Using the available data, we quantitatively study the new sum rules derived in this paper for the case of production of light quark mesons as well as mesons containing charm quarks, 
both by real photons in Sec.~\ref{sec8}, and by virtual photons in Sec.~\ref{sec9}. 
We demonstrate the intricate cancellations that must occur among the 
(pseudo-) scalar, tensor, and axial-vector mesons in order to satisfy these sum rules. In the case of production of virtual photons, 
we use these relations to provide estimates of hitherto unmeasured $\gamma^\ast \gamma$ transition form factors of tensor mesons, 
such as $f_2(1285)$ and $a_2(1320)$. 
The conclusion and outlook is given in Sec.~\ref{sec:concl}.

The Appendices contain (\ref{app:gaga}) a review of the
kinematical notations and $e^\pm + e^- \to e^\pm + e^- + X$ cross section conventions;  
(\ref{app:perturb}) expressions for the tree-level $\gamma^\ast \gamma^\ast$ cross sections for the case of 
scalar and spinor QED (Sec.~\ref{app:perturb}); 
(\ref{sec7}) general formalism for the 
$\gamma^\ast \gamma \to {\rm meson}$ 
transitions with different quantum numbers ($J^{PC}$), i.e.: 
pseudo-scalars ($0^{-+}$), scalars ($0^{++}$), axial-vectors ($1^{++}$), and tensors ($2^{++}$).

\section{Derivation of sum rules for light-light scattering}
\label{sec2}

\subsection{Forward scattering amplitudes}

In the most general case we consider the forward scattering of virtual photons on virtual photons: 
\begin{equation}
\gamma^\ast(\lambda_1, q_1) + \gamma^\ast(\lambda_2, q_2) \to 
\gamma^\ast(\lambda^\prime_1, q_1) + \gamma^\ast(\lambda^\prime_2, q_2), 
\label{eq:process1}
\end{equation} 
where $q_1$, $q_2$ are photon four-momenta, and 
$\lambda_1, \lambda_2$ ($\lambda^\prime_1, \lambda^\prime_2$) are the helicities 
of the initial (final) virtual photons, which can take on the values  $\pm 1$ (transverse polarizations) and zero (longitudinal). The total helicity in the $\gamma^\ast \gamma^\ast$ c.m. system is given by 
$\Lambda = \lambda_1 - \lambda_2 = \lambda^\prime_1 - \lambda^\prime_2$. 
To define the kinematics, we firstly introduce the photon virtualities 
$Q_1^2 = - q_1^2$, $Q_2^2 = -q_2^2$, the Mandelstam invariants: 
$s = (q_1 + q_2)^2$, $u = (q_1 - q_2)^2$,
and the following crossing-symmetric variable: 
\begin{eqnarray}
\nu \equiv \mbox{$\frac14$}(s - u) = q_1 \cdot q_2, 
\end{eqnarray}
such that
$s = 2 \nu - Q_1^2 - Q_2^2$,
$u = - 2 \nu - Q_1^2 - Q_2^2$.

The $\gamma^\ast \gamma^\ast \to \gamma^\ast \gamma^\ast$ forward scattering amplitudes, denoted as $M_{\lambda^\prime_1 \lambda^\prime_2, \lambda_1 \lambda_2}$,  are functions of $\nu$, $Q_1^2$, $Q_2^2$.  
Parity invariance ($P$) and time-reversal invariance ($T$) imply the following relations among the matrix elements with different helicities~: 
\begin{eqnarray}
P: \quad M_{\lambda^\prime_1 \lambda^\prime_2, \lambda_1 \lambda_2} &=&
M_{- \lambda^\prime_1 - \lambda^\prime_2, - \lambda_1 - \lambda_2}, 
\\
T: \quad M_{\lambda^\prime_1 \lambda^\prime_2, \lambda_1 \lambda_2} &=&
M_{\lambda_1 \lambda_2, \lambda^\prime_1 \lambda^\prime_2},
\end{eqnarray}
which leaves out only eight independent amplitudes~\cite{Budnev:1971sz}:
\begin{equation}
M_{++,++}, \, M_{+-,+-}, \, M_{++,--}, \,  
M_{00,00}, \, M_{+0,+0}, \, M_{0+,0+}, \,  M_{++,00}, \, M_{0+,-0}.
\end{equation}
 
We next look at the constraint imposed by crossing symmetry, which requires that the amplitudes for the process (\ref{eq:process1}) equal the amplitudes for the process where the photons with e.g.\ label 2 are crossed:
\begin{equation}
\gamma^\ast(\lambda_1, q_1) + \gamma^\ast(- \lambda^\prime_2, - q_2) \to 
\gamma^\ast(\lambda^\prime_1, q_1) + \gamma^\ast(- \lambda_2, - q_2).  
\label{eq:process2}
\end{equation} 
As under photon crossing $\nu \to -\nu$, one obtains 
\begin{eqnarray}
M_{\lambda^\prime_1 \lambda^\prime_2, \lambda_1 \lambda_2}(\nu, Q_1^2, Q_2^2) &=&
M_{\lambda^\prime_1 - \lambda_2, \lambda_1 - \lambda^\prime_2}(-\nu, Q_1^2, Q_2^2),
\end{eqnarray}
it becomes convenient to introduce amplitudes which are either even or odd in $\nu$ (at fixed $Q_1^2$ and $Q_2^2$). 
One easily verifies that  the following six amplitudes are {\it even} in $\nu$~:
\begin{eqnarray}
 \left( M_{++,++} + M_{+-,+-}  \right), \quad 
M_{++,--},  \quad
M_{00,00}, \quad
M_{+0,+0}, \quad
M_{0+,0+}, \quad
 \left( M_{++,00} + M_{0+,-0}  \right), 
\label{eq:even}
\end{eqnarray}
whereas the following two amplitudes are {\it odd} in $\nu$~:
\begin{eqnarray}
 \left( M_{++,++} - M_{+-,+-}  \right), \quad \quad
\left( M_{++,00} - M_{0+,-0}  \right). 
\label{eq:odd}
\end{eqnarray}

\subsection{Fusion of two virtual photons}
\label{sec3}

The optical theorem allows one to relate the absorptive part of the $\gamma^\ast \gamma^\ast \to \gamma^\ast \gamma^\ast$ forward scattering amplitudes to 
cross sections for the process $\gamma^\ast \gamma^\ast \to \mathrm{X}$ , where X stands for any possible final state.  Denoting the absorptive part as 
\begin{eqnarray}
W_{\lambda^\prime_1 \lambda^\prime_2, \lambda_1 \lambda_2} \equiv  \mathrm{Abs}\,M_{\lambda^\prime_1 \lambda^\prime_2, \lambda_1 \lambda_2}, 
\end{eqnarray} 
the optical theorem yields:
\begin{eqnarray}
W_{\lambda^\prime_1 \lambda^\prime_2, \lambda_1 \lambda_2} = 
\frac{1}{2} \int d \Gamma_\mathrm{X} (2 \pi)^4 \delta^4(q_1 + q_2 - p_\mathrm{X}) \, 
{\cal M}_{\lambda_1 \lambda_2} (q_1, q_2; p_\mathrm{X}) \, 
{\cal M}^\ast_{\lambda^\prime_1 \lambda^\prime_2} (q_1, q_2; p_\mathrm{X}), 
\label{eq:abs}
\end{eqnarray}
where ${\cal M}_{\lambda_1 \lambda_2} (q_1, q_2; p_\mathrm{X})$ denotes the invariant amplitude 
for the process
\begin{equation}
\gamma^\ast(\lambda_1, q_1) + \gamma^\ast(\lambda_2, q_2) \to \mathrm{X}(p_\mathrm{X}).
\end{equation}
As a result,  the absorptive parts are expressed in terms of eight independent 
$\gamma^\ast \gamma^\ast \to \mathrm{X}$ cross sections (see Ref.~\cite{Budnev:1974de}
for details):
\begin{subequations}
\label{eq:vcross}
\begin{eqnarray}
W_{++,++} + W_{+-,+-}  &\equiv& 2 \sqrt{X} \, \left(\sigma_0 + \sigma_2 \right) = 2 \sqrt{X} \,  \left(\sigma_\parallel + \sigma_\perp \right)
\equiv 4 \sqrt{X} \, \sigma_{TT},   \\
W_{++,++} - W_{+-,+-}   &\equiv& 2 \sqrt{X} \, \left(\sigma_0 - \sigma_2 \right) \equiv 4 \sqrt{X} \, \tau^a_{TT} ,  \\
W_{++,--} &\equiv& 2 \sqrt{X} \,  \left(\sigma_\parallel - \sigma_\perp \right) \equiv 2 \sqrt{X} \, \tau_{TT} ,  \\
W_{00,00} &\equiv& 2 \sqrt{X} \, \sigma_{LL},  \\
W_{+0,+0} &\equiv& 2 \sqrt{X} \, \sigma_{TL},  \\
W_{0+,0+} &\equiv& 2 \sqrt{X} \, \sigma_{LT},  \\
W_{++,00} + W_{0+,-0}  &\equiv& 4 \sqrt{X} \, \tau_{TL},  \\
W_{++,00} - W_{0+,-0}  &\equiv& 4 \sqrt{X} \,  \tau^a_{TL}, 
\end{eqnarray}
\end{subequations}
where the virtual photon flux factor is defined through
 \begin{equation}
X \equiv (q_1 \cdot q_2)^2 - q_1^2 q_2^2 = \nu^2 - Q_1^2 Q_2^2. 
 \end{equation}
In Eq.~(\ref{eq:vcross}), 
 $\sigma_0 (\sigma_2)$ are the $\gamma^\ast \gamma^\ast \to \mathrm{X}$ cross sections for total helicity 0 (2) respectively, 
 and $\sigma_\parallel (\sigma_\perp)$ are the  cross sections for linear photon polarizations with both photon polarization 
 directions parallel (perpendicular) to each other respectively. 
 The remaining cross sections (positive definite quantities $\sigma$) 
 involve either one transverse ($T$) and one longitudinal ($L$) photon polarization, or two longitudinal photon polarizations, with $\sigma_{LT}$ and $\sigma_{TL}$ related as~:
 \begin{equation}
\sigma_{LT}(\nu, Q_1^2, Q_2^2) = \sigma_{TL}(\nu, Q_2^2, Q_1^2). 
 \end{equation}  
The quantities $\tau_{TT}, \tau^a_{TT}, \tau_{TL}, \tau^a_{TL}$ denote interference cross sections (which are not sign-definite) 
with either both photons transverse ($TT$), or for one transverse and one longitudinal photon ($TL$),  
 where the superscript $a$ indicates the combinations which are odd in $\nu$.

\subsection{Dispersion relations}
\label{sec4}
The principle of (micro-)causality is known to translate into exact statements
about analytic properties of the scattering amplitude in the complex energy plane. 
In our case this principle translates into the statement of analyticity
of the forward $\gamma^\ast \gamma^\ast$ scattering amplitude in the
entire $\nu$ plane, except for the real axis where the branch cuts associated with
particle production are located. Assuming that the threshold for particle production is $\nu_0 > 0$, one can write down the usual dispersion relations, 
in which the amplitude is given by integrals over the non-analyticities, which in
this case are branch cuts extending from $\pm \nu_0$ to $\pm \infty$.
Finally, for amplitudes that are even or odd in $\nu$ we can write (for any fixed values of $Q_1^2, Q_2^2 > 0$):
\begin{subequations}
\begin{eqnarray}
f_{even}(\nu)  & = & \frac{2}{\pi} \int_{\nu_0}^\infty \! d \nu^\prime \frac{\nu^\prime}{\nu^{\prime \, 2} - \nu^2-i0^+} \mathrm{Abs} \, f_{even}(\nu^\prime),\\
\label{eq:dreven}
f_{odd}(\nu) & = & \frac{2 \nu}{\pi} \int_{\nu_0}^\infty \! d \nu^\prime \frac{ 1}{\nu^{\prime \, 2} - \nu^2 - i0^+} \mathrm{Abs}\, f_{odd}(\nu^\prime), 
\label{eq:drodd}
\end{eqnarray}
\end{subequations}
where $0^+$ is an infinitesimal positive number.

These dispersion relations are derived with the provision that the integrals converge.
If they do not, subtractions must be made; e.g., the once-subtracted
dispersion relation for the even amplitudes reads:
\begin{eqnarray}
f_{even}(\nu)  & = & f_{even}(0) + \frac{2\nu^2}{\pi} \int_{\nu_0}^\infty \! d \nu^\prime \frac{1}{\nu^\prime (\nu^{\prime \, 2} - \nu^2-i0^+)} \mathrm{Abs} \, f_{even}(\nu^\prime).
\label{eq:drsubt}
\end{eqnarray}
 We are thus led to examine the high-energy behavior ($\nu \to \infty$ at fixed $Q_1^2, Q_2^2$) 
of the absorptive parts given by Eq.~(\ref{eq:vcross}). In Ref.~\cite{Budnev:1971sz}, a Regge pole model assumption for the high-energy asymptotics of the 
light-by-light forward amplitudes yielded:
\begin{eqnarray}
\left(W_{++,++} + W_{+-,+-}\right), \quad   W_{+0,+0}, \quad W_{0+,0+}, \quad W_{00,00} \quad &\sim& \nu^{\alpha_P(0)}, \nonumber \\  
\left(W_{++,++} - W_{+-,+-}\right), \quad   W_{++,--} \quad &\sim& \nu^{\alpha_\pi(0)},  \\  
\left(W_{++,00} + W_{0+,-0}\right), \quad   \left(W_{++,00} -  W_{0+,-0}\right) \quad &\sim& \nu^{\alpha_\pi(0) - 1},   \nonumber
\label{eq:highen}
\end{eqnarray} 
where $\alpha_P(0) \simeq 1.08$ is the intercept of the Pomeron trajectory, and $\alpha_\pi(0) \simeq -0.014 $  is the intercept of the pion trajectory.
This means that for all the even amplitudes, except  $M_{++,00} + M_{0+,-0}$, one
can only use the subtracted dispersion relation Eq.~(\ref{eq:drsubt}). 
We therefore need the information about these amplitudes at zero
energy $\nu$. 
Anticipating the discussion of the low-energy expansion of the LbL scattering,
we can state that at $\nu =0$ these amplitudes vanish when
one of the photons is real [cf.\ Eq.~(\ref{eq:LEX})]. 
Using Eq.~(\ref{eq:vcross}) then to substitute the cross
sections in place of the absorptive parts, we obtain the following sum rules for the case of one real and one virtual photon 
(when the virtual photon flux factor becomes $X = \nu^2$):
\begin{subequations}
\begin{eqnarray}
M_{++,++} (\nu)+ M_{+-,+-}(\nu)   &=& 
\frac{4\nu^2}{\pi} \int_{\nu_0}^\infty \!\! d \nu^\prime \,  \frac{  \sigma_\parallel(\nu^\prime) + \sigma_\perp(\nu^\prime) }{\nu^{\prime \, 2} - \nu^2-i0^+}    , 
\label{eq:sr4}  \\
  M_{++,--} (\nu) &=& 
\frac{4\nu^2}{\pi} \int_{\nu_0}^\infty \!\!d \nu^\prime  \, \frac{ \sigma_\parallel(\nu^\prime) - \sigma_\perp(\nu^\prime)  }{\nu^{\prime \, 2} - \nu^2-i0^+}   , 
\label{eq:sr5}  \\
  M_{0+,0+} (\nu) &=& 
\frac{4\nu^2}{\pi} \int_{\nu_0}^\infty \!\! d \nu^\prime \,  \frac{ \sigma_{LT}(\nu^\prime) }{\nu^{\prime \, 2} - \nu^2-i0^+} , \\
  M_{+0,+0} (\nu) &=& 
\frac{4\nu^2}{\pi} \int_{\nu_0}^\infty \!\! d \nu^\prime \,  \frac{ \sigma_{TL}(\nu^\prime) }{\nu^{\prime \, 2} - \nu^2-i0^+}. 
\label{eq:sr6} 
\end{eqnarray}
We cannot write such a subtracted sum rule for $M_{00,00}$, since it
trivially vanishes when one of the photons is real. Instead, considering an 
unsubtracted dispersion relation, we find the following sum rule:
\bea
  M_{00,00} (\nu) &=& 
\frac{4}{\pi} \int_{\nu_0}^\infty \!\! d \nu^\prime \,  \frac{\nu^\prime \sqrt{X^\prime} \, \sigma_{LL}(\nu^\prime) }{\nu^{\prime \, 2} - \nu^2-i0^+},
\eea
with $X'=\nu^{\prime\, 2} - Q_1^2 Q_2^2$. At least in perturbative
QED calculations (cf. Appendix~\ref{app:perturb}), the above integral converges which seems to validate this sum rule in a 
renormalizable, perturbative field theory. 
We emphasize however that this observation is in contradiction with the expectation of non-convergence from the Regge pole model shown above. A validation of this
sum rule in non-perturbative field theory, particularly in QCD, is therefore an open issue.

For all the remaining amplitudes 
the asymptotic behavior of Eq.~(\ref{eq:highen})  
justifies the use of unsubtracted dispersion relations which, upon substituting Eq.~(\ref{eq:vcross}), lead to the following sum rules, 
valid for both photon virtual: 
\begin{eqnarray}
 M_{++,++}(\nu) - M_{+-,+-} (\nu) &=& 
\frac{4\nu}{\pi} \int_{\nu_0}^\infty \!\!d \nu^\prime  \,\frac{\sqrt{X^\prime}\,\big[ \sigma_0(\nu^\prime) - \sigma_2(\nu^\prime) \big]  }{\nu^{\prime \, 2} - \nu^2-i0^+}   , 
\label{eq:sr1} \\
 M_{++,00}(\nu) - M_{0+,-0}(\nu)  &=& 
\frac{8\nu}{\pi} \int_{\nu_0}^\infty \!\!d \nu^\prime  \,\frac{\sqrt{X^\prime} \, \tau^a_{TL} (\nu^\prime)}{\nu^{\prime \, 2} - \nu^2-i0^+}   , 
\label{eq:sr1b}\\
M_{++,00} (\nu)+ M_{0+,-0} (\nu)   &=& 
\frac{8}{\pi} \int_{\nu_0}^\infty \!\!d \nu^\prime \, \frac{\nu^\prime \sqrt{X^\prime}\, \tau_{TL} (\nu^\prime) }{\nu^{\prime \, 2} - \nu^2-i0^+} ,
\label{eq:sr2} 
\end{eqnarray}
where the dependence on virtualities $Q_1^2$, $Q_2^2$ is tacitly  assumed.
\label{eq:sumrules} 
\end{subequations}

The above sum rules, relating all the forward $\gamma^\ast \gamma^\ast$ elastic scattering
amplitudes to the energy integrals of  the $\gamma^\ast \gamma^\ast$ fusion cross sections,
should hold for any space-like photon virtualities in the unsubtracted cases, and for
one of the virtualities equal to zero in the subtracted cases. In the following
we examine the low-energy expansion of these sum rules. 

\subsection{Low-energy expansion via effective Lagrangian}
\label{sec:sr}

To obtain more specific relations from the sum rules established in Eq.~(\ref{eq:sumrules}), we parametrize the low-energy (small $\nu$) 
behavior of the $\gamma^\ast \gamma^\ast \to \gamma^\ast \gamma^\ast$ forward scattering amplitudes $M$.  
At lowest order in the energy, the self-interactions of the electromagnetic field  are described by an effective Lagrangian (of fourth order in the photon energy and/or momentum, and fourth order in the electromagnetic field):
\begin{equation}
\mathcal{L}^{(8)} = c_1 (F_{\mu\nu}F^{\mu\nu})^2 + c_2 (F_{\mu\nu}\tilde F^{\mu\nu})^2,
\label{EHL} 
\end{equation}
where $F_{\mu\nu} = \partial_\mu A_\nu - \partial_\nu A_\mu$, $\tilde F^{\mu\nu} = 
\varepsilon^{\mu\nu\alpha\beta} \partial_\alpha A_\beta$, 
and where $c_1, c_2$ are two low-energy constants (LECs) which contain the structure dependent information. 
It is often referred to as Euler-Heisenberg Lagrangian due to 
the seminal work~\cite{Heisenberg:1935qt}. 

At the next order in energy, one considers the terms involving two derivatives on the field tensors, corresponding with the sixth order in the photon energy and/or momentum. Writing down all such dimension-ten operators and reducing their number using the antisymmetry of the field tensors, the Bianchi identities, as well as adding or removing total derivative terms, 
we find that there are 6 independent terms at that order, which we choose as~:
\begin{eqnarray}
\mathcal{L}^{(10)}&=&
c_{3} (\pd_{\a} F_{\m\n}) (\pd^{\a} F^{\l\n}) F_{\l\r}F^{\m\r}+
c_4 (\pd_{\a} F_{\m\n}) (\pd^{\a} F^{\m\n}) F_{\l\r}F^{\l\r}
\nonumber \\
&+ & c_5 (\pd^{\a} F_{{\a}\n}) (\pd_\b F^{\b\n}) F_{\l\r}F^{\l\r} 
+c_{6} (\pd_{\a}\pd^{\a} F_{\m\n}) F^{\l\n}F_{\l\r}F^{\m\r}\nonumber \\
&+&
c_7 ( \pd_{\a}\pd^{\a} F_{\m\n}) F^{\m\n}F_{\l\r}F^{\l\r}+
c_{8} (\pd^{\a}  F_{\a\m}) (\pd_{\b} F^{\b\l})  F_{\r\l}F^{\r\m},
\end{eqnarray}
where $c_3, \ldots, c_8$ are the new LECs arising at this order. 
Only $c_3$ and $c_4$ appear in the case of real photons. 

We can now specify the low-energy limit of the light-by-light scattering
amplitudes in terms of the LECs describing the low-energy self-interactions of
the electromagnetic field:
\begin{subequations}
\label{eq:LEX}
\begin{eqnarray}
M_{++,++} + M_{+-,+-}  &=& Q_1^2Q_2^2 \left[ 64(c_1-c_2)+ 4(Q_1^2+Q_2^2)(-c_3-8c_4-4c_5+8c_7-c_8)+{\cal O}(Q^4)  \right] \nn\\
& + & 8 \nu^2\left[ 8 (c_1 + c_2) + \left (Q_1^2 + Q_2^2  \right) (-c_3+3c_{6}+4c_{7})  + {\cal O}(Q^4)  \right] + {\cal O}(\nu^4), 
\label{eq:lex4} \\
M_{++,--}  &=& Q_1^2Q_2^2 \left[ 64c_2+4(Q_1^2+Q_2^2)(-c_3+2c_6-c_8)+{\cal O}(Q^4)  \right] \nn\\
&+& 8\nu^2 \left[ 8 (c_1 - c_2) + \left (Q_1^2 + Q_2^2\right) ( c_{6}+4c_7 )  
+{\cal O}(Q^4) \right] + {\cal O}(\nu^4), 
\label{eq:lex5} \\
M_{0+,0+}   &=& Q_1^2Q_2^2\left[-32 c_1+4Q_1^2 c_8+4(Q_1^2+Q_2^2)(c_3+4c_4+2c_5-2c_6-4c_7) +{\cal O}(Q^4)  \right] \nn\\
&+& \nu^2 \left[  
-4Q_1^2 c_{8}  +{\cal O}(Q^4)  \right] + {\cal O}(\nu^4), \\
M_{+0,+0} &=&  Q_1^2Q_2^2\left[-32 c_1+4Q_2^2 c_8+4(Q_1^2+Q_2^2)(c_3+4c_4+2c_5-2c_6-4c_7) +{\cal O}(Q^4)  \right] \nn\\
&+& \nu^2 \left[  
-4Q_2^2 c_{8}  +{\cal O}(Q^4)  \right] + {\cal O}(\nu^4), \\
M_{00,00}   &=& Q_1^2 Q_2^2 \left[ 96 c_1 +4(Q_1^2+Q_2^2)(-2c_3-4c_4-2c_5+6c_6+12c_7-c_8) + {\cal O}(Q^4) \right] \nn\\
&+& {\cal O}(\nu^2),
\label{eq:lex7}  \\
M_{++,++} - M_{+-,+-}  &=& 8 \nu Q_1^2 Q_2^2 \left[  -c_3 - 4 c_{5}  +  c_{8}  +  {\cal O}(Q^2) \right] + \n^3\left[-64c_4 +  {\cal O}(Q^2) \right]+ {\cal O}(\nu^5), 
\label{eq:lex1} \\
M_{++,00} - M_{0+,-0} &=& \nu Q_1 Q_2 \left[ - 64 c_1 
+ \left(Q_1^2 + Q_2^2 
\right) (4c_3-16c_{6}-32c_{7}+4c_{8}) + {\cal O}(Q^4) \right] \nn\\
&+& {\cal O}(\nu^3), 
\label{eq:lex2} \\
M_{++,00} + M_{0+,-0}    &=& Q_1^3 Q_2^3 \left[ 4c_5-12c_{8}  + {\cal O}(Q^2)  \right] +4\n^2 Q_1Q_2  \left[ 2 c_3+16c_4+4c_5+c_8+  {\cal O}(Q^2)  \right]
\nn\\
&+& {\cal O}(\nu^4). \label{eq:lex3} 
\label{eq:lex6} 
\end{eqnarray}
\end{subequations}
These expressions can be treated as a simultaneous expansion in $\nu$ and
the virtualities $Q_i^2$ of the {\em lhs} of the sum rules Eq.~(\ref{eq:sumrules}).
Concerning the $Q$ dependence, it is
important that the leading in $\nu$ term, in any of the amplitudes, is proportional to 
$Q_1 Q_2$ and hence vanishes for at least one real photon.
The latter statement is valid for any values of virtualities, not just when they are small.
For example, let us show  for the amplitude $( M_{++,++} - M_{+-,+-} )$ its leading term in $\nu$ is proportional to the combination $Q_1^2Q_2^2$, to all orders in $Q_1$ and $Q_2$. 

Since all photons are transversely polarized the only non-vanishing structures involving polarization vectors of photons $\varepsilon ( \lambda_i)$ are their mutual scalar products 
$\varepsilon (\lambda_i) \cdot\varepsilon (\lambda_j)$. Due to gauge invariance, the electromagnetic fields enter the Lagrangian 
through the field tensor $F_{\mu\nu}$, which contributes to the amplitude as $q_\mu\varepsilon_\nu-q_\nu\varepsilon_\mu$. Thus an arbitrary term in the effective Lagrangian contributes to $( M_{++,++} - M_{+-,+-} )$ as:
\begin{equation}
\label{eq:qT}
M_{++,++} - M_{+-,+-} \sim q_1^\mu q_2^\nu q_1^\lambda q_2^\rho T_{\mu\nu\lambda\rho},
\end{equation}
where the tensor $T_{\mu\nu\lambda\rho}$ is constructed from four-vectors $q_{i}$ and the metric tensor. Since this amplitude is odd with respect to $\nu$, it is required to be proportional to at least $\nu^1$. Assuming that one factor $\nu$ comes from contraction of two of 
the $q$'s in Eq.~(\ref{eq:qT}), we are left with $q_1^\mu q_2^\nu$. Now, if we suppose that $q_1$ is contracted with $q_2$ we obtain an extra power of $\nu$, and such an amplitude vanishes when taking the limit $\nu\rightarrow 0$. Thus, both $q_1$ and $q_2$ must be contracted with another $q_1$ and $q_2$ respectively, giving a global factor $Q_1^2Q_2^2$.

We are now in position to examine the sum rules in Eq.~(\ref{eq:sumrules}) order by order in $\nu$. For this we expand the {\em rhs} of Eq.~(\ref{eq:sumrules}) using 
$1/(\nu^{\prime \, 2} - \nu^2) = 1/\nu^{\prime \, 2} + \nu^2/\nu^{\prime \, 4} + {\cal O}(\nu^4)$. As the result we obtain 
from Eqs.~(\ref{eq:sr1},\ref{eq:sr1b},\ref{eq:sr2}) the following set of super-convergence relations, valid for at least one real photon (e.g., $Q_1\geq 0$,  $Q_2^2 = 0$): 
\begin{subequations}
\begin{eqnarray}
0 &=& \int\limits_{s_0}^\infty d s  \frac{ 1 }{(s + Q_1^2)} \, \tau_{TT}^a (s, Q_1^2, 0), 
\label{s0rule1} \\
0 &=& \int\limits_{s_0}^\infty d s  \, \frac{1}{(s + Q_1^2)^2} 
\left[ \sigma_\parallel + \sigma_{LT} + \frac{(s + Q_1^2)}{Q_1 Q_2} \tau^a_{TL} 
\right]_{Q_2^2 = 0}, 
\label{s0rule4} \\
0 &=& \int\limits_{s_0}^\infty d s  \, \left[ \frac{\tau_{TL} (s, Q_1^2, Q_2^2) }{Q_1 Q_2} 
\right]_{Q_2^2 = 0}. 
\label{s0rule6}
\end{eqnarray}
\end{subequations}
and the following set of sum rules for the LECs of the dimension-8 (Euler-Heisenberg)
Lagrangian, valid when both  photons
are quasi-real:
\begin{subequations}
\begin{eqnarray}
 c_1 & = & \frac{1}{8 \pi }\int\limits_{s_0}^{\infty} \frac{d s}{s^2}
 \,\sigma_\parallel (s,0,0),
 \label{s1rule1} \\ 
 &=& - \frac{1}{ 8 \pi} \int\limits_{s_0}^\infty 
 \frac{ds }{s}  \, \left[ \frac{\tau^a_{TL} (s, Q_1^2, Q_2^2) }{Q_1 Q_2} 
\right]_{Q_1^2 = Q_2^2 = 0}, 
\label{s0rule3} \\
&=& \frac{1}{ 8 \pi} \int\limits_{s_0}^\infty d s\, \left[ \frac{\si_{LL} (s, Q_1^2, Q_2^2) }{Q_1^2 Q_2^2} 
\right]_{Q_1^2 = Q_2^2 = 0}, 
\label{s1rule8} \\
c_2 & = & \frac{1}{8 \pi }\int\limits_{s_0}^{\infty} \frac{d s}{s^2}\,  \sigma_\perp(s,0,0),
 \label{s0rule2}
\end{eqnarray}
\end{subequations}
where $s_0 = 2 \nu_0 - Q_1^2 - Q_2^2$. 
We emphasize again that, 
unlike the other sum rules, the sum rule of Eq.~(\ref{s1rule8}) is only shown to hold in perturbative field theory. 

There are as well the sum rules for the LECs of the dimension-10 Lagrangian, most notably:
\begin{eqnarray}
c_4 &=& - \frac{1}{ 4 \pi}\int\limits_{s_0}^\infty \frac{ ds }{s^3} \, \tau_{TT}^a (s, 0, 0), 
\label{s1rule9} 
\end{eqnarray}
but presently they are of far lesser importance and we do not write them out 
here explicitly.

Let us remark again that the relation of Eq.~(\ref{s0rule1}), obtained by
combining Eqs.~(\ref{eq:sr1}) and (\ref{eq:lex1}), 
is essentially a GDH sum rule for the photon target, see
\cite{Gerasimov:1973ja,Brodsky:1995fj,Roy:1974fz}. For large virtuality $Q_1^2$, 
it leads to the sum rule for the photon structure function $g_1^\gamma$
\cite{Bass:1998bw}:
$\int_0^1 d x g_1^\gamma(x, Q^2) = 0$.

The sum rules in Eqs. (\ref{s1rule1}) and (\ref{s0rule2}), first established in \cite{Pascalutsa:2010sj},  are obtained by combining Eqs.~(\ref{eq:sr4}) with (\ref{eq:lex4}) and Eqs.~(\ref{eq:sr5}) with (\ref{eq:lex5}), respectively. 
All the other relations presented above are new.
In the following section we verify these sum rules in QED at leading order
in the fine-structure constant $\a$.

\section{Sum rules in perturbation theory}
\label{pert}

We will subsequently discuss a pair production in scalar QED (e.g., Born approximation to $\gamma^\ast \gamma^\ast \to \pi^+ \pi^-$)  
and in spinor QED ($\gamma^\ast \gamma^\ast \to q \bar q$ where $q$ stands 
for a charged lepton or a quark).

\subsection{Scalar QED}

\begin{figure}[h]
\includegraphics[width =10cm]{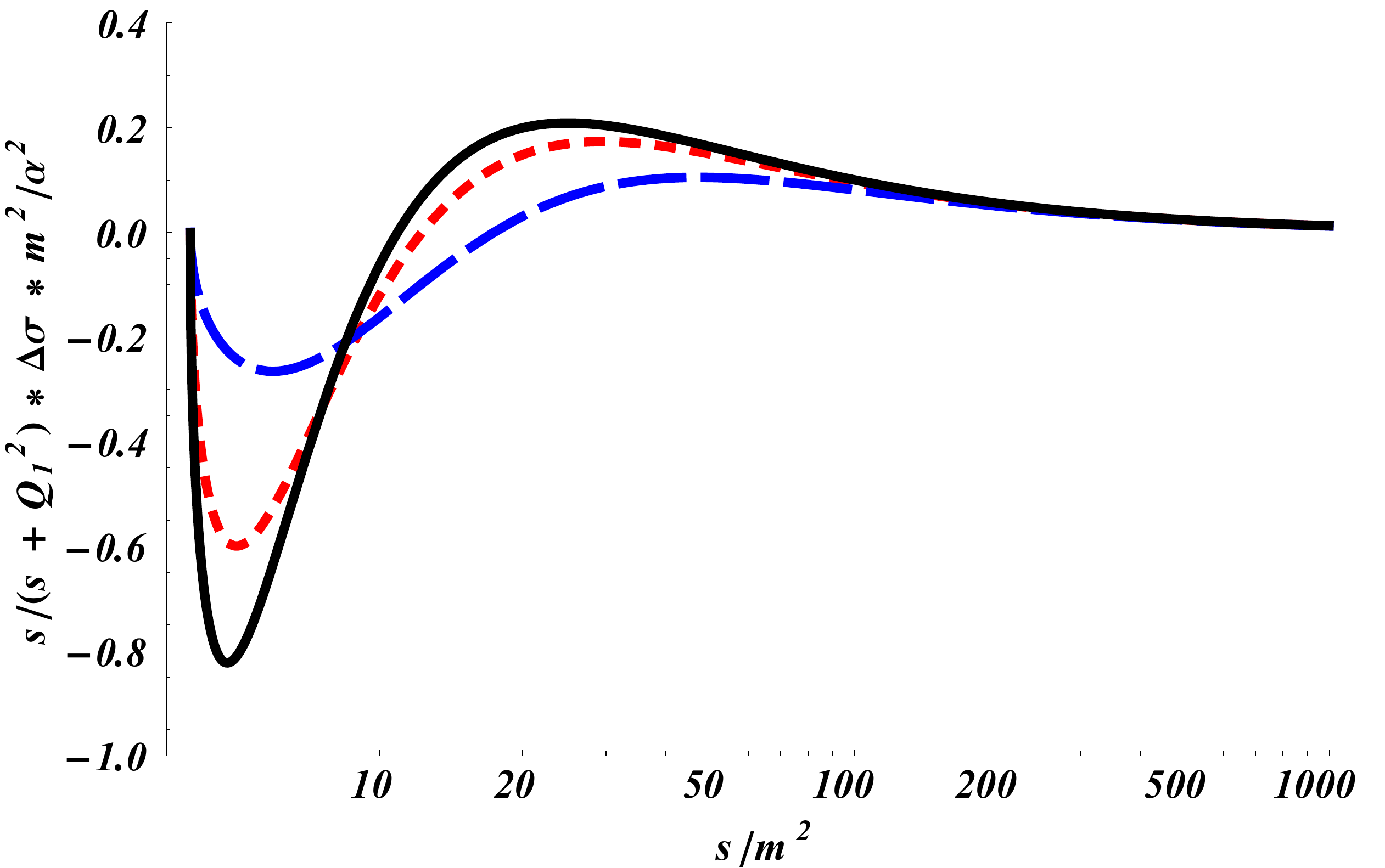}
\caption{The $\gamma^\ast \gamma \to \pi^+ \pi^-$ tree level result (scalar QED) for the 
integrand in the  $\Delta \sigma \equiv \sigma_2 - \sigma_0$ sum rule  of Eq.~(\ref{s0rule1}), 
multiplied by $s$, where one of the photons is real. 
The different curves are for different virtualities for the other photon~:
$Q_1^2 = 0$ (solid black curve), $Q_1^2 = m^2$ (short-dashed red curve), $Q_1^2 = 5 m^2$ (long-dashed blue curve).  }
\label{fig:scalarsr1}
\end{figure}

\begin{figure}[h]
\includegraphics[width =10cm]{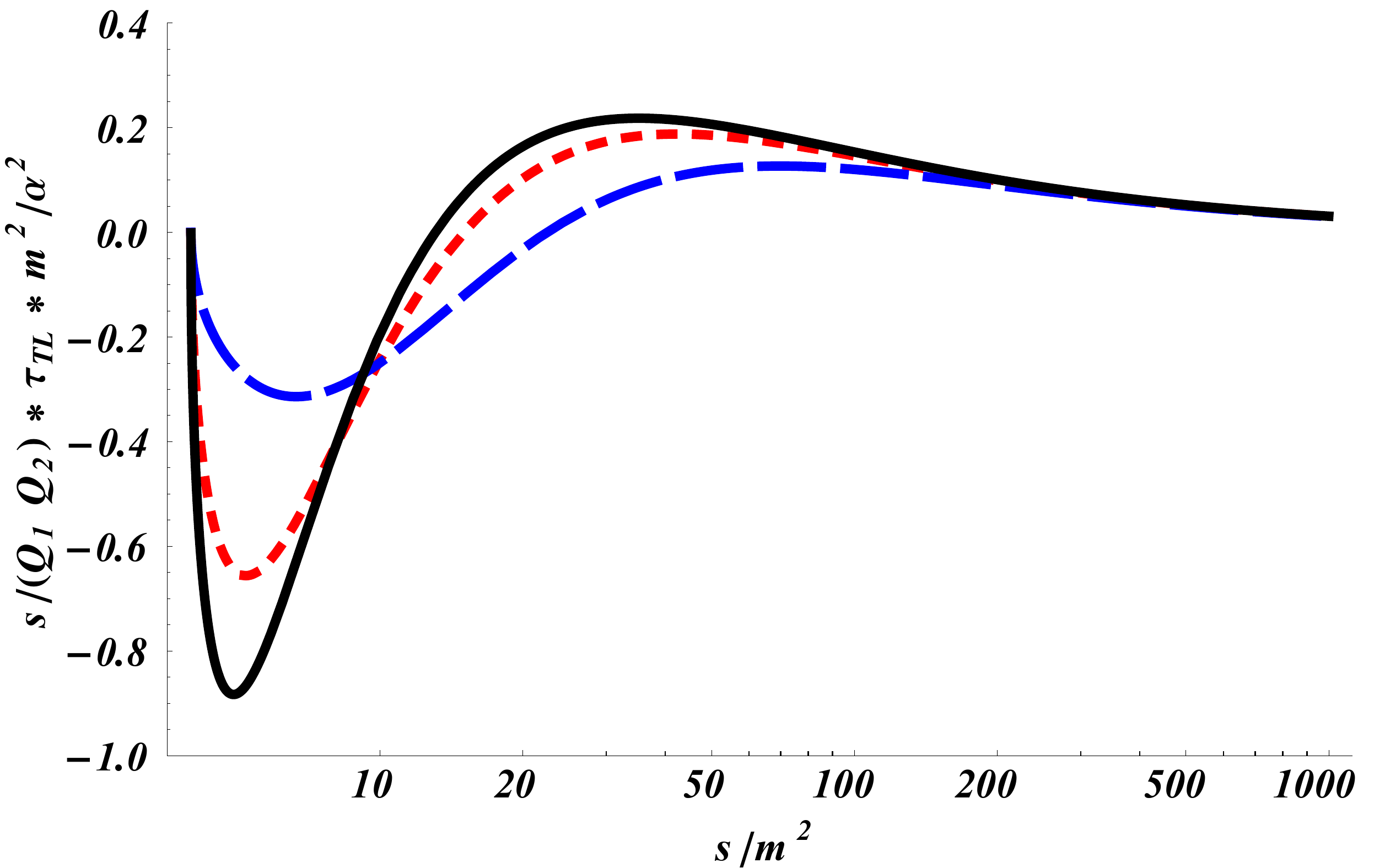}
\caption{The $\gamma^\ast \gamma \to \pi^+ \pi^-$ tree level result (scalar QED) for the integrand 
in the $\tau_{TL}$ sum rule of Eq.~(\ref{s0rule6}) multiplied by $s$, 
where one of the photons is quasi-real. The different curves are for different virtualities for the other photon~:
$Q_1^2 = 0$ (solid black curve), $Q_1^2 = m^2$ (short-dashed red curve), $Q_1^2 = 5 m^2$ (long-dashed blue curve).  }
\label{fig:scalarsr2}
\end{figure}

\begin{figure}[h]
\includegraphics[width =10cm]{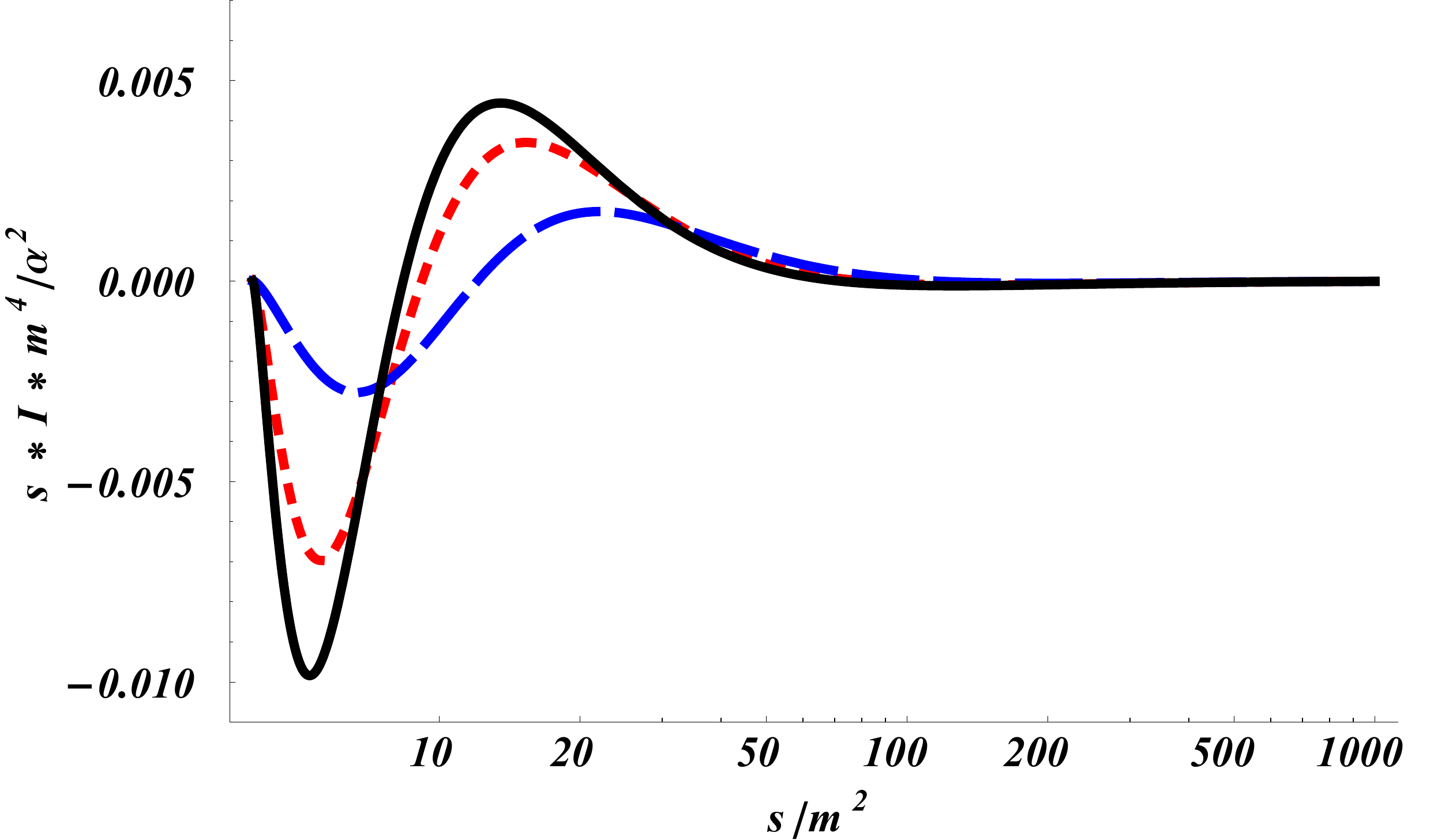}
\caption{The $\gamma^\ast \gamma \to \pi^+ \pi^-$ tree level result (scalar QED) for the integrand 
$I$ in the sum rule of Eq.~(\ref{s0rule4}) multiplied by $s$, with $I$ given by Eq.~(\ref{eq:intI}),   
where one of the photons is quasi-real. The different curves are for different virtualities for the other photon~:
$Q_1^2 = 0$ (solid black curve), $Q_1^2 = m^2$ (short-dashed red curve), $Q_1^2 = 5 m^2$ (long-dashed blue curve).  }
\label{fig:scalarsr3}
\end{figure}

The response functions for the case of scalar QED at lowest order in the electromagnetic coupling can be found in Appendix~\ref{app:scalar}.
We firstly study the three sum rules of Eqs.~(\ref{s0rule1}, \ref{s0rule4}, \ref{s0rule6}) for the case of one real or quasi-real photon ($Q_2^2 \to 0$) and for arbitrary space-like virtuality ($Q_1^2 \geq 0$) of the other photon.  
To better see the cancellation which must take place in these sum rules between contributions at low and higher energies, 
we show the integrands of the three sum rules in Figs.~\ref{fig:scalarsr1}, \ref{fig:scalarsr2}, \ref{fig:scalarsr3} multiplied 
by $s$. In this way, when plotted logarithmically, one can clearly see how the low and high energy contributions 
cancel each other. For the sum rule of Eq.~(\ref{s0rule4}), we denote the integrand as~:
\begin{eqnarray}
I = \frac{1}{(s + Q_1^2)^2} 
\left[ \sigma_\parallel + \sigma_{LT} + \frac{(s + Q_1^2)}{Q_1 Q_2} \tau^a_{TL} 
\right]_{Q_2^2 = 0}. 
\label{eq:intI}
\end{eqnarray}
All three sum rules of Eqs.~(\ref{s0rule1}, \ref{s0rule4}, \ref{s0rule6}) are exactly verified in scalar QED for arbitrary 
space-like values of $Q_1^2$. One notices from Figs.~\ref{fig:scalarsr1}, \ref{fig:scalarsr2}, \ref{fig:scalarsr3} that for larger values 
of $Q_1^2$ the zero crossing of the integrands shifts to larger values of $s$, requiring higher energy contributions for the cancellation to take place. For the helicity difference sum rule of Eq.~(\ref{s0rule1}), one notices that at low energies 
$\sigma_0$ dominates while with increasing energies $\sigma_2$ overtakes.

Besides exactly verifying the sum rules which integrate to zero, we can also use the above derived sum rules to study the 
low-energy coefficients for light-by-light scattering in scalar QED. 
Using Eqs.~(\ref{s1rule1}, \ref{s0rule2}), we obtain for the tree-level contributions to the lowest order coefficients $c_1$ and $c_2$ in scalar QED:
\begin{eqnarray}
c_1 = \frac{\alpha^2}{m^4} \frac{7}{1440}, 
\quad \quad \quad
c_2 = \frac{\alpha^2}{m^4} \frac{1}{1440}.
\end{eqnarray}

\subsection{Spinor QED}

\begin{figure}[h]
\includegraphics[width =10cm]{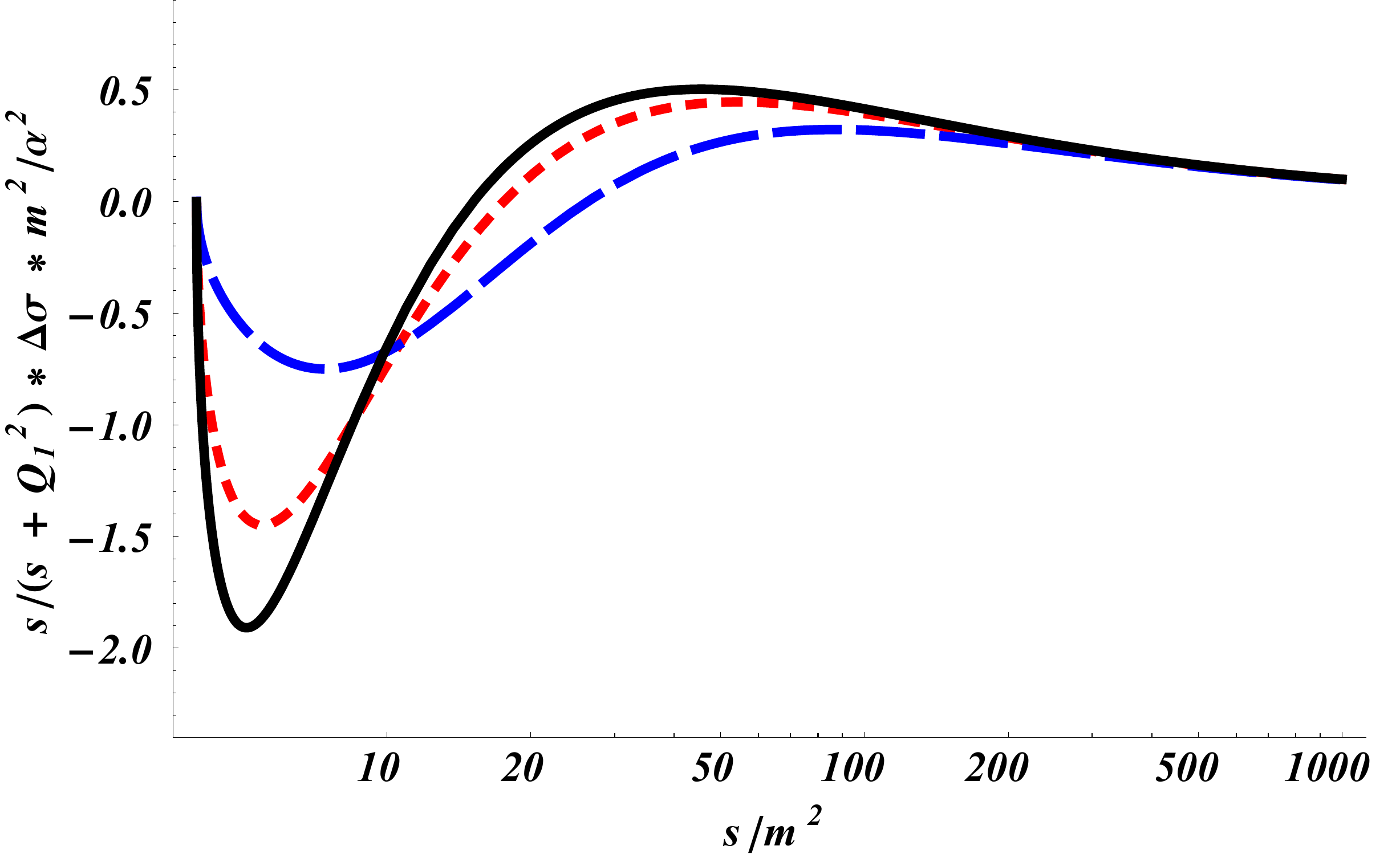}
\caption{The $\gamma^\ast \gamma \to e^+ e^-$ tree level result (spinor QED) for the 
integrand in the  $\Delta \sigma \equiv \sigma_2 - \sigma_0$ sum rule  of Eq.~(\ref{s0rule1}), 
multiplied by $s$, where one of the photons is real. 
The different curves are for different virtualities for the other photon~:
$Q_1^2 = 0$ (solid black curve), $Q_1^2 = m^2$ (short-dashed red curve), $Q_1^2 = 5 m^2$ (long-dashed blue curve).  }
\label{fig:spinorsr1}
\end{figure}

\begin{figure}[h]
\includegraphics[width =10cm]{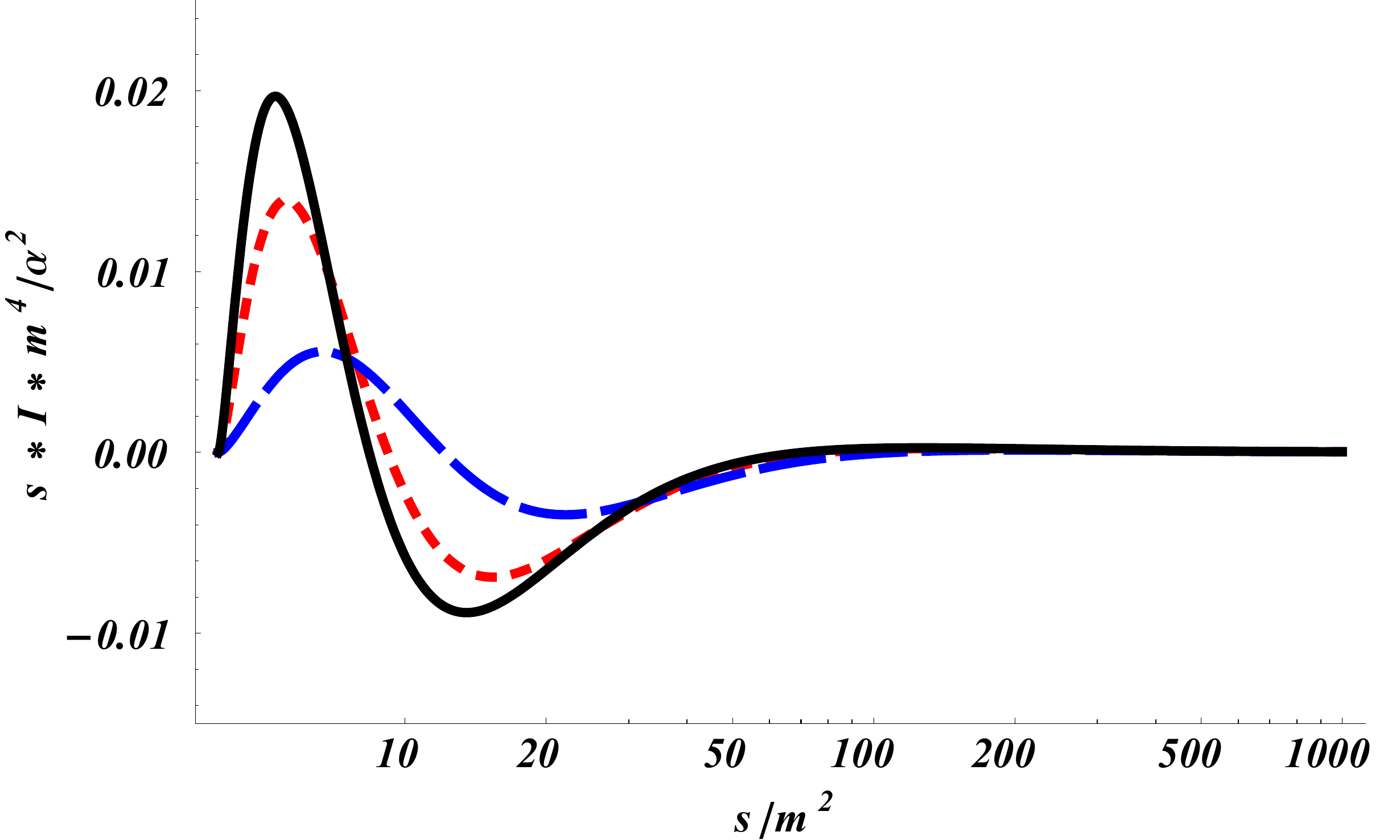}
\caption{The $\gamma^\ast \gamma \to e^+ e^-$ tree level result (spinor QED) for the integrand 
$I$ in the sum rule of Eq.~(\ref{s0rule4}) multiplied by $s$, with $I$ given by Eq.~(\ref{eq:intI}),   
where one of the photons is quasi-real. The different curves are for different virtualities for the other photon~:
$Q_1^2 = 0$ (solid black curve), $Q_1^2 = m^2$ (short-dashed red curve), $Q_1^2 = 5 m^2$ (long-dashed blue curve).  }
\label{fig:spinorsr3}
\end{figure}

The response functions for the case of spinor QED at lowest order in the electromagnetic coupling can be found in Appendix~\ref{app:spinor}.
We again study the three sum rules of Eqs.~(\ref{s0rule1}, \ref{s0rule4}, \ref{s0rule6}) for the case of one real or quasi-real photon ($Q_2^2 \to 0$) for different space-like virtualities of the other photon.  
As the tree level contribution to $\tau_{TL}$ in spinor QED vanishes for one quasi-real photon, one notices 
that the sum rule of Eq.~(\ref{s0rule6}) is trivially satisfied.  
For the sum rules involving the helicity difference of Eq.~(\ref{s0rule1}), and 
involving the integrand $I$ of Eq.~(\ref{eq:intI}),  
we show the corresponding integrands multiplied by $s$ 
in Figs.~\ref{fig:spinorsr1}, \ref{fig:spinorsr3} for the case of one real or quasi-real photon and for different virtualities of the 
other photon. 
We again verify that the sum rules involve an exact cancellation between 
low and high energy contributions. 

Using Eqs.~(\ref{s1rule1}, \ref{s0rule2}), we obtain for the tree-level contributions to the lowest order coefficients $c_1$ and $c_2$ 
for light-by-light scattering in spinor QED~:
\begin{eqnarray}
c_1 = \frac{\alpha^2}{m^4} \frac{1}{90}, 
\quad \quad \quad
c_2 = \frac{\alpha^2}{m^4} \frac{7}{360}.
\end{eqnarray}

In these case we also were able to verify the sum rule in Eq. (\ref{s1rule9}), yielding
\beq
c_4 = - \frac{\a^2}{m^6} \frac{1}{315}, 
\eeq
in agreement with the result obtained in Ref.~\cite{Dicus:1997ax} for the low-energy photon-photon scattering.

A more detailed study of the LbL sum rules in field theory, including 
loop effects, production of vector bosons, etc., is the subject of
our forthcoming publication \cite{Vlad}.

\section{Meson production in $\gamma\gamma$ collision}

In the previous section, the sum rules of Eqs.~(\ref{s0rule1}, \ref{s0rule4}, \ref{s0rule6}) integrating to zero 
have been shown to hold exactly in perturbative calculations (e.g., in QED or QCD in the perturbative regime). 
However as their derivation is general, their realization 
in QCD, in its non-perturbative regime, 
allows to gain insight in the $\gamma^\ast \gamma \to {\rm hadrons}$ cross-sections. 
This was illustrated in Ref.~\cite{Pascalutsa:2010sj} for the sum rule of Eq.~(\ref{s0rule1}).  In the remainder of this paper,  
we will elaborate on the discussion of Ref.~\cite{Pascalutsa:2010sj} and extend it to the other sum rules presented above. 
The required non-perturbative input for the absorptive parts of the sum rules are the $\gamma^\ast \gamma \to {\rm hadrons}$ response functions.  
In this paper, we will perform a first analysis by estimating the hadronic contributions to these response functions by the corresponding $\gamma^\ast \gamma^\ast \to M$ (with $M$ a meson) production processes, which are described in terms of the $\gamma^\ast \gamma^\ast \to M$ transition form factors. 

In Appendix \ref{sec7} we detail the formalism and the available data for the
$\gamma^\ast \gamma^\ast \to M$ transition FFs, and 
 successively discuss the $C$-even
pseudo-scalar ($J^{PC} = 0^{-+}$), scalar ($J^{PC} = 0^{++}$), axial-vector ($J^{PC} = 1^{++}$), 
and tensor ($J^{PC} = 2^{++}$) mesons.

\subsection{Real photons}
\label{sec8}

We first consider the helicity sum rule of Eq.~(\ref{s0rule1}) with two real photons producing a meson, 
as well as the sum rules of Eq.~(\ref{s0rule2}) for the mesonic contributions to the low-energy constants $c_1$ and $c_2$ 
describing the forward light-by-light scattering amplitude.  
When producing mesons, the sum rules will hold separately for states of given intrinsic quantum numbers. 
Therefore, we will separately study the sum rule contributions for light quark isovector mesons (Table~\ref{table_isovector}), 
for light quark isoscalar mesons (Table~\ref{table_isoscalar}), as well as $c \bar c$ mesons (Table~\ref{table_ccbar}).  
For the isoscalar mesons, one could in principle separate the contributions according to singlet or octet states (or alternatively 
according to $(u \bar u + d \bar d)/\sqrt{2}$ or $s \bar s$ states). The corresponding mesons 
involve mixings however which complicate such separation, as this mixing is not known well enough for some of the states. 
We will postpone such a separation for a future work and add all isoscalar meson contributions in the present work.  

The pseudo-scalar mesons contribute to the helicity-0 cross section only, given by Eq.~(\ref{eq:pscross}). 
The corresponding contributions to the helicity sum rule of Eq.~(\ref{s0rule1})  as well as the $c_1$ and $c_2$ sum rules  
are shown for the $\pi^0$ in Table~\ref{table_isoscalar}, for the $\eta, \eta^\prime$ in Table~\ref{table_isoscalar}, 
and for the $\eta_c(1S)$ state in Table~\ref{table_ccbar}.
 
Besides the pseudo-scalar mesons, also scalar mesons can only contribute to $\sigma_0$. We show the contributions of the 
$a_0(980)$ in Table~\ref{table_isoscalar}, for the $f_0(980)$ and $f_0^\prime(1370)$ in Table~\ref{table_isoscalar}, 
and for the $\chi_{c0}(1P)$ state in Table~\ref{table_ccbar}. For the scalar mesons, only the $f_0^\prime(1370)$ state gives 
a sizable contribution due to its large $2 \gamma$ decay width.  
    
For the helicity sum rule, one notices that in order to 
compensate the large negative contribution from the pseudo-scalar mesons, and to lesser extent from the scalar meson states,
an equal strength is required in the helicity-2 cross section, $\sigma_2$.
For light quark mesons, the dominant feature of the helicity-2 cross section in the resonance region  
arises from the multiplet of tensor mesons $f_2(1270)$, $a_2(1320)$, and $f_2^\prime(1525)$. For $c \bar c$ tensor mesons, the 
dominant tensor contribution is given by the $\chi_{c2}(1P)$ state. 

Measurements at various $e^+ e^-$ colliders, notably the recent high statistics measurements 
by the BELLE Collaboration of the $\gamma \gamma$ cross sections to 
$\pi^+ \pi^-$, $\pi^0 \pi^0$, $\eta \pi^0$, and $K^+ K^-$ channels~\cite{bellepipi,belle2pi0,belleetapi0} 
have allowed to accurately establish their parameters. 
For the light quark mesons, the experimental analyses of decay angular distributions 
have found~\cite{Pennington:2008xd} that the tensor mesons are produced predominantly (around 95\% or more) 
in a state of helicity $\Lambda = 2$. We will therefore assume in all of the following analyses that 
$\Gamma_{\gamma \gamma}\left({\cal T}(\Lambda = 0) \right) \approx 0$, and that 
$\Gamma_{\gamma \gamma}\left({\cal T}(\Lambda = 2) \right) \approx \Gamma_{\gamma \gamma}({\cal T})$ 
in Tables~\ref{table_isovector}, \ref{table_isoscalar}, \ref{table_ccbar}. 
We show all tensor meson contributions to the helicity difference sum rule as well as the $c_1, c_2$ sum rules for which the $2 \gamma$ decay widths are known. 

For the isovector meson contributions to the helicity sum rule, 
shown in Table~\ref{table_isovector}, we conclude  that the lowest isovector tensor meson composed of light quarks, $a_2(1320)$, compensates to around 70\% the contribution of the $\pi^0$, which is entirely governed by the chiral anomaly. 
For the isoscalar states composed of light quarks,  the cancellation is even more remarkable:  the sum of  $f_2(1270)$ and $f_2^\prime(1525)$, within the experimental accuracy, entirely compensates the combined contribution of the $\eta$ and $\eta^\prime$ mesons.   

\begin{table}[h]
{\centering \begin{tabular}{|c|c|c|c|c|c|}
\hline
& $m_M$  & $\Gamma_{\gamma \gamma} $   &   $\int \frac{ds}{s}\;  (\sigma_2 - \sigma_0) $ & $c_1$ & $c_2$  \\
&  [MeV] &  [keV] &   [nb]  & [$10^{-4}$\, GeV$^{-4}$] &  [$10^{-4}$\, GeV$^{-4}$] \\
\hline 
\hline
\quad $\pi^0$ \quad  & \quad $134.9766  \pm 0.0006$ \quad  & \quad  $(7.8 \pm 0.5) \times 10^{-3}$ \quad  &  $ -195 \pm 13$ & \quad 0 \quad 
&  \quad $10.94 \pm 0.70$ \quad \\
\hline
\quad $a_0(980)$  \quad &  \quad $980 \pm 20$  \quad  & \quad  $0.3 \pm 0.1$ \quad  &  $ -20 \pm 8 $ 
& \quad $0.021 \pm 0.007$ \quad & \quad 0 \quad  \\
\hline
\quad $a_2 (1320)$ \quad &  \quad $1318.3 \pm 0.6$  \quad  & \quad  $1.00 \pm 0.06$ \quad   & \quad  $134 \pm 8$ \quad 
& \quad $0.039 \pm 0.002$  \quad & \quad $0.039 \pm 0.002$  \quad  \\
\quad $a_2 (1700)$ \quad &  \quad $1732 \pm 16$  \quad  & \quad  $0.30 \pm 0.05$ \quad  &   $18 \pm 3$   
& \quad $0.003 \pm 0.001$  \quad & \quad $0.003 \pm 0.001$  \quad  \\
\hline
\hline
Sum &  & & $ \quad -63 \pm 17$ \quad & \quad $0.06 \pm 0.01$ \quad & \quad $10.98 \pm 0.70$ \quad \\
\hline
\end{tabular}\par}
\caption{$\gamma \gamma$ sum rule contributions of the light quark isovector mesons based on the present PDG values~\cite{PDG} of the meson masses ($m_M$) and their $2 \gamma$ decay widths $\Gamma_{\gamma \gamma}$. Fourth column: $\sigma_2 - \sigma_0$ sum rule of Eq.~(\ref{s0rule1}).
Fifth, sixth columns: $c_1, c_2$ sum rules of Eqs.~(\ref{s1rule1}, \ref{s0rule2}) respectively. }
\label{table_isovector}
\end{table}

\begin{table}[h]
{\centering \begin{tabular}{|c|c|c|c|c|c|}
\hline
& $m_M$  & $\Gamma_{\gamma \gamma} $  &   $\int \frac{ds}{s}\;  (\sigma_2 - \sigma_0)  $  & $c_1$ & $c_2$ \\
&  [MeV] &  [keV] &   [nb]  & [$10^{-4}$GeV$^{-4}$] &  [$10^{-4}$GeV$^{-4}$] \\
\hline 
\hline
\quad $\eta$ \quad & \quad  $547.853 \pm 0.024$ \quad   & \quad  $0.510 \pm 0.026$ \quad   &  \quad $-191 \pm 10 $ \quad  
& \quad 0 \quad & \quad $0.65 \pm 0.03$ \quad  \\
\quad $\eta^\prime$ \quad &\quad  $957.78 \pm 0.06$ \quad   & \quad  $4.29 \pm 0.14$ \quad  & \quad $ -300 \pm 10$  \quad 
& \quad 0 \quad & \quad $0.33 \pm 0.01$ \quad   \\
\hline
\quad $f_0(980)$ \quad  & \quad $980 \pm 10$ \quad   & \quad  $0.29 \pm 0.07$  \quad   &  $ -19 \pm 5 $  & 
\quad $0.020 \pm 0.005$ \quad & \quad 0 \quad \\
\quad $f_0^\prime(1370)$  \quad & \quad $1200 - 1500 $  \quad  & \quad  $3.8 \pm 1.5$ \quad &  $ -91 \pm 36 $ 
& \quad $0.049 \pm 0.019$ \quad & \quad 0 \quad  \\
\hline
\quad $f_2 (1270)$ \quad & \quad $1275.1 \pm 1.2 $ \quad   & \quad  $3.03 \pm 0.35$  \quad  &  \quad $449 \pm 52$  \quad 
&\quad $0.141 \pm  0.016$ \quad  &\quad $0.141 \pm  0.016$ \quad   \\
\quad $f_2^\prime (1525)$ \quad & \quad $1525 \pm 5$  \quad  & \quad  $0.081 \pm 0.009$ \quad &  \quad $7 \pm 1$ \quad  
&\quad $0.002 \pm  0.000$ \quad  &\quad $0.002 \pm  0.000$ \quad  \\
\quad $f_2 (1565)$ \quad & \quad $1562 \pm 13 $ \quad   & \quad  $0.70 \pm 0.14$  \quad &  \quad $56 \pm 11$   \quad 
&\quad $0.012 \pm  0.002$ \quad  &\quad $0.012 \pm  0.002$ \quad  \\
\hline
\hline
Sum &  & & \quad $-89 \pm 66$ \quad & \quad $0.22 \pm 0.03$ \quad & \quad $1.14 \pm 0.04$ \quad \\
\hline
\end{tabular}\par}
\caption{$\gamma \gamma$ sum rule contributions of the light quark isoscalar mesons based on the present PDG values~\cite{PDG} of the meson masses ($m_M$) and their $2 \gamma$ decay widths $\Gamma_{\gamma \gamma}$. Fourth column: $\sigma_2 - \sigma_0$ sum rule of Eq.~(\ref{s0rule1}).  
Fifth, sixth columns: $c_1, c_2$ sum rules of Eqs.~(\ref{s1rule1}, \ref{s0rule2}) respectively. }
\label{table_isoscalar}
\end{table}

\begin{table}[h]
{\centering \begin{tabular}{|c|c|c|c|c|c|}
\hline
& $m_M$  & $\Gamma_{\gamma \gamma} $  &   $\int \frac{ds}{s}\;  (\sigma_2 - \sigma_0) $ & $c_1$ & $c_2$ \\
&  [MeV] &  [keV] &   [nb] & [$10^{-7}$GeV$^{-4}$] &  [$10^{-7}$GeV$^{-4}$]  \\
\hline 
\hline
\quad $\eta_c(1S)$ \quad  & \quad  $2980.3 \pm 1.2$ \quad   & \quad  $6.7 \pm 0.9$ \quad   &  \quad $\quad  -15.6 \pm 2.1$ \quad   
& \quad 0 \quad &  \quad $1.79 \pm 0.24$ \quad \\
\hline
\quad $\chi_{c0}(1P)$  \quad  &  \quad $3414.75 \pm 0.31$  \quad  & \quad  $2.32 \pm 0.13$ \quad &  $ \quad -3.6 \pm 0.2 \quad $ 
& \quad $0.31 \pm 0.02$ \quad & \quad 0 \quad  \\
\hline
\quad $\chi_{c2}(1P)$ \quad &  \quad $3556.2 \pm 0.09$  \quad  & \quad  $0.50 \pm 0.06$ \quad   & $ \quad 3.4 \pm 0.4$   \quad  
& \quad $0.14 \pm 0.02$ \quad & \quad $0.14 \pm 0.02$ \quad \\
\hline
\hline
\quad Sum resonances \quad &  & & \quad $ -15.8 \pm 2.1$ \quad & \quad $0.49 \pm 0.03$ \quad & \quad $1.97 \pm 0.24$ \quad \\
\hline
\hline
\quad duality estimate \quad &  & & & &  \\
\quad continuum ($\sqrt{s} \geq 2m_D$) \quad &  & &  \quad $15.1$ \quad & & \\
\hline
\hline
\quad resonances + continuum \quad &  & &  \quad $-0.7 \pm 2.1$ \quad & & \\
\hline
\end{tabular}\par}
\caption{$\gamma \gamma$ sum rule contributions of the lowest $c \bar c$ mesons based on the present PDG values~\cite{PDG} of the meson masses ($m_M$) and their $2 \gamma$ decay widths $\Gamma_{\gamma \gamma}$.  
Fourth column: the $\sigma_2 - \sigma_0$ sum rule of Eq.~(\ref{s0rule1}), for which we also show the duality estimate 
of Eq.~(\ref{eq:duality3}) 
for the continuum contribution above $D \bar D$ threshold, as well as the sum of  resonances and continuum 
contributions.  
Fifth, sixth columns: $c_1, c_2$ sum rules of Eqs.~(\ref{s1rule1}, \ref{s0rule2}) respectively. }
\label{table_ccbar}
\end{table}

\begin{figure}[h]
\includegraphics[width =11cm]{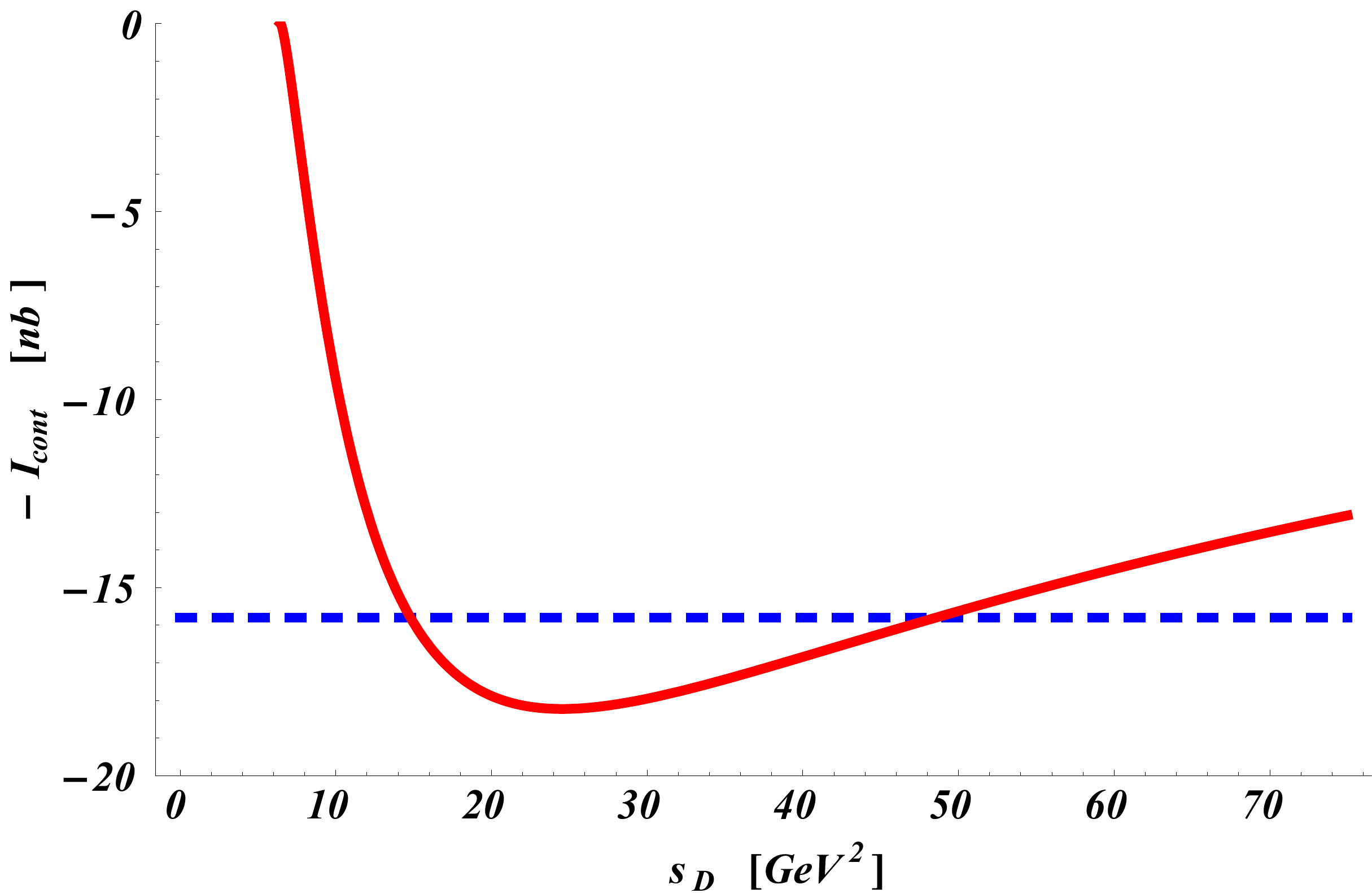}
\caption{Solid (red) curve: 
duality estimate for the negative of the continuum contribution of Eq.~(\ref{eq:duality3}) to the helicity difference sum rule for 
charm quarks as function of the integration limit $s_D$, which represents the threshold for charmed meson production 
($D \bar D$ threshold). For reference, the dashed (blue) horizontal curve indicates the sum of the 
$\eta_c(1S), \chi_{c0}(1P)$, and $\chi_{c2}(1P)$ resonance contributions to the $\sigma_2 - \sigma_0$ sum rule, 
as listed in Table~\ref{table_ccbar}. 
The intersection between both curves near the physical $D \bar D$ threshold, $s_D \approx 14$~GeV$^2$ indicates a perfect cancellation between these resonance contributions and the duality estimate for the continuum contribution.
 }
\label{fig:duality}
\end{figure}

For the $c \bar c$ states, one notices that the known strength in the tensor channel from the $\chi_{c2}(1P)$ state 
only compensates about 20\% of the strength arising from the $\eta_c(1S)$ and $\chi_{c0}(1P)$ states. We can 
however expect a sizable contribution to this sum rule from states above the nearby $D \bar D$ threshold, which we 
denote by $s_D = 4 m_D^2 \approx 14$~GeV$^2$, using the $D$-meson mass $m_D \approx 1.87$~GeV.
So far, the helicity cross sections have not been measured above $D \bar D$ threshold. To estimate this 
continuum contribution to the helicity sum rule, which we denote by $I_{cont}$, 
we use a quark-hadron duality argument~\cite{Novikov:1977dq} , which amounts to replacing the 
integral of the helicity difference cross section 
for the $\gamma \gamma  \to X$ process (with $X$ any hadronic final state containing charm quarks)
by the corresponding integral of the helicity difference cross section for the 
perturbative $\gamma \gamma \to c \bar c$ process~:
\begin{eqnarray}
I_{cont} \equiv \int \limits_{s_D}^{\infty} \, ds \, \frac{1}{s} \left[ \sigma_2 - \sigma_0 \right] (\gamma \gamma \to X) 
\approx \int \limits_{s_D}^{\infty} \, ds \, \frac{1}{s} \left[ \sigma_2 - \sigma_0 \right] (\gamma \gamma \to c \bar c),  
\label{eq:duality1}
\end{eqnarray}
where the perturbative cross section is given in Appendix~\ref{app:spinor}. 
The duality expressed by the approximate equality in Eq.~(\ref{eq:duality1}) is meant to hold in a global sense, i.e. after integration over the 
energy of the helicity difference cross section above the threshold $s_D$.  
As we have verified in Section~\ref{pert} that the perturbative cross section satisfies the helicity sum rule exactly, 
i.e.
\begin{eqnarray}
0 = \int \limits_{4 m_c^2}^{\infty} \, ds \, \frac{1}{s} \left[ \sigma_2 - \sigma_0 \right] (\gamma \gamma \to c \bar c),  
\end{eqnarray}
with $m_c$ the charm quark mass, we can re-express Eq.~(\ref{eq:duality1}) as~:
\begin{eqnarray}
I_{cont} 
\approx - \int \limits_{4 m_c^2}^{s_D} \, ds \, \frac{1}{s} \left[ \sigma_2 - \sigma_0 \right] (\gamma \gamma \to c \bar c). 
\label{eq:duality2}
\end{eqnarray}
Using Eq.~(\ref{eq:spinheldiff}) for the $\gamma \gamma \to c \bar c$ helicity difference cross section, we finally obtain:
\begin{eqnarray}
I_{cont} 
\approx - 8 \pi \, \alpha^2 \, \int \limits_{4 m_c^2}^{s_D} \, ds \, \frac{1}{s^2} 
\left\{ -3 \,  \sqrt{1 - \frac{4 m_c^2}{s}} \,
+ 2 \, \ln \left( \frac{\sqrt{s}}{2 m_c} \left[ 1 + \sqrt{1 - \frac{4 m_c^2}{s}} \right] \right) 
\right\}.
\label{eq:duality3}
\end{eqnarray}
Using the PDG value $m_c \approx 1.27$~GeV \cite{PDG}, we show the duality estimate for $-I_{cont}$ in Fig.~\ref{fig:duality}, 
as function of the integration limit $s_D$ (solid red curve). 
Using the physical value of the $D \bar D$ threshold, $s_D \approx 14$~GeV$^2$, 
we obtain: $I_{cont} \approx 15.1$~nb. We notice that within the experimental uncertainty, this fully cancels the 
sum of the $\eta_c(1S), \chi_{c0}(1P)$, and $\chi_{c2}(1P)$ resonance contributions to the 
$\sigma_2 - \sigma_0$ sum rule, as is shown in Table~\ref{table_ccbar}. 
This cancellation quantitatively illustrates the interplay between resonances with hidden charm ($c \bar c$ states) 
and production of charmed mesons in order to satisfy the sum rule. It will be interesting to further test this experimentally by 
measuring the $\gamma \gamma$ production cross sections above $D \bar D$ threshold, where a plethora of new 
states (so-called $XYZ$ states) have been found in recent years, see e.g. Ref.~\cite{Godfrey:2008nc} for a review. 

We have also computed the meson contributions to the forward light-by-light scattering coefficients $c_1$ and $c_2$ (fifth and sixth columns respectively in Tables~\ref{table_isovector}, \ref{table_isoscalar}, \ref{table_ccbar}). The dimensionality of these coefficients requires them 
to scale with the meson mass $m_M$ as $1/m_M^4$. Therefore, the higher mass mesons contribute very insignificantly to these coefficients. 
One notes that the coefficient $c_1$, which involves the cross section $\sigma_\parallel$, does not receive any contributions from pseudo-scalar mesons, and is dominated by the tensor mesons $a_2(1320)$ and $f_2(1270)$, with smaller contributions from the scalar states around 1 GeV. On the other hand, the  coefficient $c_2$, which involves the cross section $\sigma_\perp$, is totally dominated by the  contributions from pseudo-scalar mesons, especially the light $\pi^0$, with contributions of $\eta$ and $\eta^\prime$ at the 10\% level of the $\pi^0$ contribution.

\subsection{Virtual photons}
\label{sec9}

We next discuss the sum rule of Eq.~(\ref{s0rule4}) when both photons are quasi-real. 
One immediately observes that pseudo-scalar mesons do not contribute to this sum rule. However scalar, axial-vector and tensor mesons will contribute to this sum rule. The sum rule will therefore require a cancellation mechanisms between 
scalar, axial-vector and tensor mesons, which we will study subsequently.  
According to Eq.~(\ref{eq:scross}), scalar mesons (with mass $m_S$) can only contribute to the $\sigma_\parallel$ term in the sum rule, 
and their contribution is given by:
\begin{eqnarray}
\int ds \, \frac{1}{s^2} \left[ \sigma_\parallel \right]_{Q_1^2 = Q_2^2 = 0} 
= 16 \pi^2 \, \frac{\Gamma_{\gamma \gamma} ({\cal S})}{m_S^5}.
\label{sr2:s}
\end{eqnarray}

In contrast, Eq.~(\ref{eq:across}) shows that axial-vector mesons  (with mass $m_A$) can only 
contribute to the $\tau^a_{TL}$ term in the sum rule as:
\begin{eqnarray}
\int ds \, \frac{1}{s} \left[\frac{\tau^a_{TL}}{Q_1 Q_2} \right]_{Q_1^2 = Q_2^2 = 0}  =  - 8 \pi^2 \, 
\frac{3 \, \tilde \Gamma_{\gamma \gamma} ({\cal A})}{m_A^5},
\label{sr2:a}
\end{eqnarray}
where we introduced the equivalent $2 \gamma$ decay width $\tilde \Gamma_{\gamma \gamma} ({\cal A})$ of Eq.~(\ref{a2gwidth}). 

The tensor mesons in general contribute to both terms of the sum rule of Eq.~(\ref{s0rule4}). 
For the $\sigma_\parallel$ contribution, we will use the experimental observation that  light tensor mesons are produced predominantly (around 95 \% or more) in a state of helicity $\Lambda = 2$, as discussed above. Neglecting therefore the much smaller $\sigma_0$ term, we obtain
from Eq.~(\ref{eq:tcross}):
\begin{eqnarray}
\int ds \, \frac{1}{s^2} \left[ \sigma_\parallel \right]_{Q_1^2 = Q_2^2 = 0} 
= \int ds \, \frac{1}{s^2} \frac{1}{2} \left[ \sigma_2 \right]_{Q_1^2 = Q_2^2 = 0} 
= 8 \pi^2 \, \frac{5 \, \Gamma_{\gamma \gamma} ({\cal T})}{m_T^5},
\label{sr2:t1}
\end{eqnarray}
with tensor meson mass $m_T$. 
For the $\tau^a_{TL}$ contribution to the sum rule of Eq.~(\ref{s0rule4}), one sees 
from Eq.~(\ref{eq:tcross}) 
that it involves a helicity-1 amplitude for tensor meson production by quasi-real photons, 
which unfortunately is not known experimentally for any tensor meson. 
It is reasonable to assume that for quasi-real photons this amplitude is much smaller than the 
helicity-2 amplitude which is known to dominate in the real photon limit.  
We will therefore neglect the helicity-1 contribution in the following analysis. 

One notes from Eqs.~(\ref{sr2:s}, \ref{sr2:a}, \ref{sr2:t1}) that only axial-vector mesons give a negative contribution to 
the sum rule of Eq.~(\ref{s0rule4}), 
whereas scalar and tensor mesons contribute positively. As the sum rule has to integrate to zero, one therefore obtains a cancellation 
mechanism between axial-vector mesons on one hand, and scalar and tensor mesons on the other. 
In Table~\ref{table_isoscalar_sr2}, we show the contributions of the lowest lying scalar, axial-vector and tensor mesons, for which the 
$2 \gamma$ widths are known experimentally. 
One sees from Table~\ref{table_isoscalar_sr2} that the two lowest lying axial-vector mesons $f_1 (1285)$ 
and $f_1(1420)$ are entirely cancelled, within error bars, by the   
contribution of the dominant tensor meson $f_2 (1270)$. Using the experimentally known $2 \gamma$ widths, the 
deviation of the (zero) sum rule value is at the $2 \sigma$ level, which hints at a moderate contribution of 
either another higher mass axial-vector meson state or a non-resonant contribution with axial-vector quantum numbers. 

\begin{table}[h]
{\centering \begin{tabular}{|c|c|c|c|c|c|}
\hline
& $m_M$  & $\Gamma_{\gamma \gamma} $  &   $\int  \frac{ds}{s^2} \, \sigma_\parallel(s) $  & $\int ds  \;  
\left[ \frac{1}{s} \frac{\tau^a_{TL}}{Q_1 Q_2} \right]_{Q_i^2  = 0}$   
& $\int ds \; \left[ \frac{1}{s^2} \sigma_\parallel  + \frac{1}{s} \frac{\tau^a_{TL}}{Q_1 Q_2} \right]_{Q_i^2  = 0}$   \\
&  [MeV] &  [keV] &   [nb / GeV$^2$]  & [nb / GeV$^2$] &  [nb / GeV$^2$] \\
\hline 
\hline
\quad $f_1 (1285)$  \quad & \quad $1281.8 \pm 0.6$ \quad   & \quad  $3.5 \pm 0.8 $  \quad  & \quad  $0$ \quad 
& $ -93 \pm 21 $ &  $ -93 \pm 21 $ \\
\quad $f_1 (1420)$ \quad  & \quad $1426.4 \pm 0.9$  \quad  & \quad  $ 3.2 \pm 0.9 $ \quad  & \quad  $0$ \quad 
& $ -50 \pm 14 $ &  $ -50 \pm 14$  \\
\hline
\hline
\quad $f_0(980)$ \quad  & \quad $980 \pm 10$ \quad   & \quad  $0.29 \pm 0.07$  \quad   &  $ 20 \pm  5$  & 
\quad $0$ \quad & \quad $ 20 \pm  5$  \quad \\
\quad $f_0^\prime(1370)$  \quad & \quad $1200 - 1500 $  \quad  & \quad  $3.8 \pm 1.5$ \quad &  $  48 \pm 19 $ 
& \quad $0$ \quad & \quad $  48 \pm 19 $ \quad  \\
\hline
\quad $f_2 (1270)$ \quad & \quad $1275.1 \pm 1.2 $ \quad   & \quad  $3.03 \pm 0.35$  \quad  &  \quad $ 138 \pm 16$  \quad 
&\quad $\gtrsim 0$ \quad  &\quad $ 138 \pm 16 $ \quad   \\
\quad $f_2^\prime (1525)$ \quad & \quad $1525 \pm 5$  \quad  & \quad  $0.081 \pm 0.009$ \quad &  \quad $ 1.5 \pm 0.2$ \quad  
&\quad $\gtrsim 0$ \quad  &\quad $ 1.5 \pm  0.2$ \quad  \\
\quad $f_2 (1565)$ \quad & \quad $1562 \pm 13 $ \quad   & \quad  $0.70 \pm 0.14$  \quad &  \quad $ 12 \pm 2$   \quad 
&\quad $\gtrsim 0$ \quad  &\quad $ 12 \pm 2 $ \quad  \\
\hline
\hline
Sum &  & &  &  & \quad $ 76 \pm 36$ \quad \\
\hline
\end{tabular}\par}
\caption{Light isoscalar meson contributions to the sum rule of Eq.~(\ref{s0rule4}) based on the present PDG values~\cite{PDG} of the meson masses ($m_M$) and their $2 \gamma$ decay widths $\Gamma_{\gamma \gamma}$. 
For the axial-vector mesons, we quote the equivalent $2 \gamma$ decay width $\tilde \Gamma_{\gamma \gamma}$ of 
Table~\ref{tab_ax}. 
Fourth column: $\sigma_\parallel$ contribution, 
fifth column: $\tau^a_{TL}$ contribution, 
sixth column: total contribution to the sum rule of Eq.~(\ref{s0rule4}). 
}
\label{table_isoscalar_sr2}
\end{table}

At finite $Q_1^2$, for $Q_2^2 = 0$, the three sum rules of Eqs.~(\ref{s0rule1},  \ref{s0rule4}, \ref{s0rule6}) 
imply relations between the transition form factors for the contributing mesons. 
To date, experimental results for the $\gamma^\ast \gamma \to {\rm meson}$  FFs only exist for the 
pseudo-scalar mesons $\pi^0, \eta, \eta^\prime$, and $\eta_c(1S)$, as well as for the axial-vector mesons $f_1(1285)$, and $f_1(1420)$. 
For other mesons, in particular the tensor mesons, the corresponding form factors  
still wait to be extracted. We have seen from Table~\ref{table_isoscalar} that for real photons the dominant contributions 
to the helicity sum rule of Eq.~(\ref{s0rule1}) come from $\eta, \eta^\prime$, and $f_2(1270)$ mesons, where 
the $f_2(1270)$ contribution cancels to 90\% the contribution from the $\eta$ and $\eta^\prime$ mesons. We will therefore use the 
corresponding sum rule of Eq.~(\ref{s0rule1}) at finite $Q_1^2$ to estimate the $\gamma^\ast \gamma \to f_2(1270)$ 
helicity-2 FF from the measured $\eta$ and $\eta^\prime$ FFs, given by Eq.~(\ref{eq:psffmono}). 
Assuming that the helicity sum rule of Eq.~(\ref{s0rule1}) is saturated by the $\eta$, $\eta^\prime$, and $f_2(1270)$ mesons, we then obtain:
\begin{eqnarray}
\frac{5 \, \Gamma_{\gamma \gamma}(f_2)}{m^3_{f_2}} \, \left[ \frac{T^{(2)}_{f_2}(Q_1^2, 0)}{T^{(2)}_{f_2}(0, 0)} \right]^2 
\simeq c_\eta \, \frac{1}{\left( 1 + Q_1^2 / \Lambda^2_\eta \right)^2} 
+ c_{\eta^\prime} \, \frac{1}{\left( 1 + Q_1^2 / \Lambda^2_{\eta^\prime} \right)^2} ,
\label{eq:sr1ff}
\end{eqnarray}
where we have introduced the shorthand notation: 
\begin{eqnarray}
c_P \equiv \frac{\Gamma_{\gamma \gamma}({\cal P})}{m_P^3}.
\end{eqnarray}
For $Q_1^2 = 0$,  the $f_2(1270)$ meson contribution cancels to 90\% the $\eta + \eta^\prime$ contributions to the helicity sum rule. 
We can therefore use 
\begin{eqnarray}
\frac{5 \, \Gamma_{\gamma \gamma}(f_2)}{m^3_{f_2}} \simeq c_\eta + c_\eta^\prime, 
\end{eqnarray}
which allows us to express Eq.~(\ref{eq:sr1ff}) as: 
\begin{eqnarray}
\frac{T^{(2)}_{f_2}(Q_1^2,0)}{T^{(2)}_{f_2}(0,0)} \simeq 
\left[ \frac{c_\eta}{c_\eta + c_{\eta^\prime}} \, \frac{1}{\left( 1 + Q_1^2 / \Lambda^2_\eta \right)^2} 
+ \frac{c_{\eta^\prime}}{c_\eta + c_{\eta^\prime}} \, \frac{1}{\left( 1 + Q_1^2 / \Lambda^2_{\eta^\prime} \right)^2} 
\right]^{1/2}.
\label{eq:tensorffsr1}
\end{eqnarray}

We can obtain a second estimate for the $T^{(2)}$ FF for the $f_2(1270)$ meson from the sum rule of Eq.~(\ref{s0rule4}).
We have seen from Table~\ref{table_isoscalar_sr2} that for quasi-real photons the dominant contributions 
to this sum rule come from $f_1(1285), f_1(1420)$, and $f_2(1270)$ mesons, where 
the $f_2(1270)$ contribution cancels to 95 \% the contribution from the $f_1(1285)$ and $f_1(1420)$ mesons. 
We can then also use the corresponding sum rule of Eq.~(\ref{s0rule4}) at finite $Q_1^2$ to estimate the $\gamma^\ast \gamma \to f_2(1270)$ 
helicity-2 FF from the measured $f_1(1285)$ and $f_1(1420)$ FFs, using Eqs.~(\ref{eq:axffcahn}, \ref{eq:axffcahnexp}). 
Assuming that the helicity sum rule of Eq.~(\ref{s0rule4}) is saturated by the $f_1(1285)$, $f_1(1420)$, and $f_2(1270)$ mesons, 
which we denote by $f_1, f_1^\prime$, and $f_2$ respectively, 
and retaining only the supposedly dominant $\Lambda = 2$ FF for the tensor mesons, we obtain:
\begin{eqnarray}
\frac{5 \, \Gamma_{\gamma \gamma}(f_2)}{m^5_{f_2}} \, \frac{1}{\left( 1 + \frac{Q_1^2}{m^2_{f_2}} \right)} \, 
\left[ \frac{T^{(2)}_{f_2}(Q_1^2, 0)}{T^{(2)}_{f_2}(0, 0)} \right]^2 
\simeq c_{f_1} \, \frac{1}{\left( 1 + Q_1^2 / \Lambda^2_{f_1} \right)^4} 
+ c_{f_1^\prime} \, \frac{1}{\left( 1 + Q_1^2 / \Lambda^2_{f_1^\prime} \right)^4} ,
\label{eq:sr2ff}
\end{eqnarray}
where 
\begin{eqnarray}
c_A \equiv \frac{3 \, \tilde \Gamma_{\gamma \gamma}({\cal A})}{m_A^5}.  
\end{eqnarray}
For $Q_1^2 = 0$,  the $f_2(1270)$ meson contribution cancels to 95\% the $f_1(1285) + f_1(1420)$ contributions to the 
sum rule of Eq.~(\ref{s0rule4}), which implies:
\begin{eqnarray}
\frac{5 \, \Gamma_{\gamma \gamma}(f_2)}{m^5_{f_2}} \simeq c_{f_1} + c_{f_1^\prime}. 
\end{eqnarray}
This allows to obtain a second estimate for the $T^{(2)}$ FF for the $f_2(1270)$ meson as:
\begin{eqnarray}
\frac{T^{(2)}_{f_2}(Q_1^2,0)}{T^{(2)}_{f_2}(0,0)} \simeq \left( 1 + \frac{Q_1^2}{m^2_{f_2}} \right)^{1/2} \, 
\left[ \frac{c_{f_1}}{c_{f_1} + c_{f_1^\prime}} \, \frac{1}{\left( 1 + Q_1^2 /  \Lambda^2_{f_1} \right)^4} 
+ \frac{c_{f_1^\prime}}{c_{f_1} + c_{f_1^\prime}} \, \frac{1}{\left( 1 + Q_1^2 / \Lambda^2_{f^\prime_1} \right)^4} 
\right]^{1/2}.
\label{eq:tensorffsr2}
\end{eqnarray}

\begin{figure}[h]
\includegraphics[width =12cm]{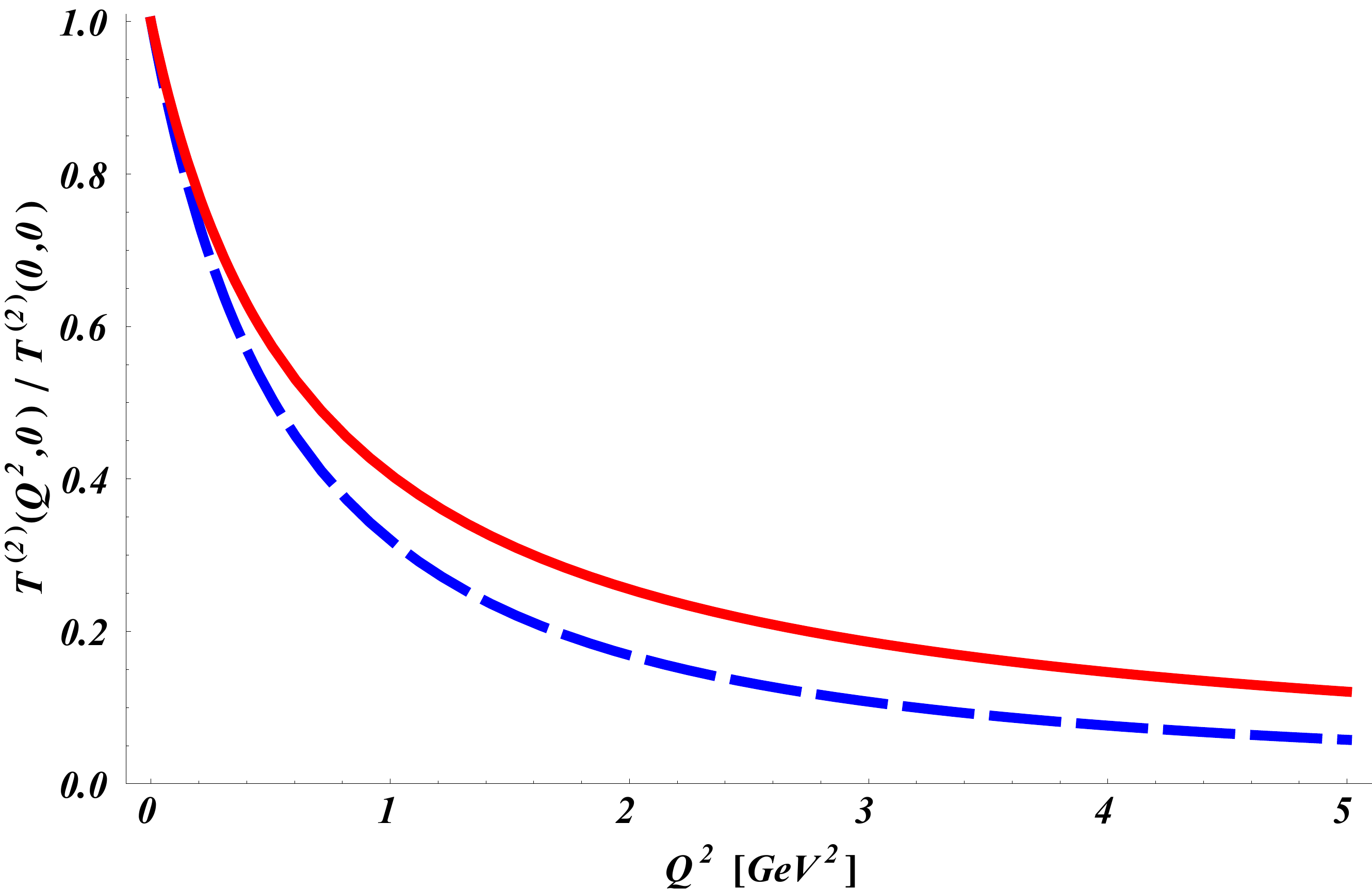}
\caption{Sum rule estimates for the form factor $T^{(2)}(Q^2,0) / T^{(2)}(0,0)$ with helicity $\Lambda = 2$ for the tensor meson $f_2(1270)$. 
Red solid curve: sum rule estimate from Eq.~(\ref{eq:tensorffsr1}), using the experimental input from the $\eta$ and $\eta^\prime$ FFs. 
Blue dashed curve: sum rule estimate from Eq.~(\ref{eq:tensorffsr2}), using the experimental input from the $f_1(1285)$ and $f_1(1420)$ FFs. 
 }
\label{fig:tensorff}
\end{figure}

In Fig.~\ref{fig:tensorff} we show the two sum rule estimates of Eqs.~(\ref{eq:tensorffsr1}) and (\ref{eq:tensorffsr2}) 
for the FF $T^{(2)}$ for the tensor meson $f_2(1270)$ using the known experimental information for either $\eta, \eta^\prime$ in 
Eq.~(\ref{eq:tensorffsr1}), or $f_1(1285), f_1(1420)$ in Eq.~(\ref{eq:tensorffsr2}). When taking the ratio of both estimates, one 
sees that it is larger than 80\% below 1 GeV$^2$ and around 65\% around $Q^2 = 2$~GeV$^2$. 
It will be interesting to confront these estimates with a direct measurement of the $T^{(2)}$ FF for the $f_2(1270)$ tensor meson.

In an analogous way, we can provide an estimate for the $a_2(1320)$ FF from the $\pi^0$ FF. 
We have seen from Table~\ref{table_isovector} that $\pi^0$ and $a_2(1320)$ provide the dominant isovector contributions to the 
helicity sum rule of Eq.~(\ref{s0rule1}), where 
the $a_2(1320)$ contribution cancels to 70\% the contribution from the $\pi^0$. We can therefore use the sum rule of 
Eq.~(\ref{s0rule1}) for one virtual photon to estimate the helicity-two FF $T^{(2)}$ for the $a_2(1320)$ meson in terms of the 
$\pi^0$ FF, given by Eq.~(\ref{eq:psffmono}), as: 
\begin{eqnarray}
\frac{T^{(2)}_{a_2}(Q_1^2,0)}{T^{(2)}_{a_2}(0,0)} \simeq 
 \frac{1}{\left( 1 + Q_1^2 / \Lambda^2_\pi \right)} .
\label{eq:ffa2}
\end{eqnarray}
As empirically the $\gamma^\ast \gamma \to \pi^0$ FF is the best known meson transition FF, it will be interesting to test the above  prediction for the $a_2(1320)$ FF experimentally.

\section{Conclusions and outlook}
\label{sec:concl}

We have studied the forward light-by-light scattering and derived three sum rules which involve 
energy weighted integrals of $\gamma^\ast \gamma$ fusion cross sections, 
measurable at $e^+ e^-$ colliders, which integrate to zero (super-convergence relations):
\begin{eqnarray}
0 &=& \int\limits_{s_0}^\infty d s  \frac{1}{(s + Q_1^2)}  \left[ \sigma_0 - \sigma_2 \right]_{Q_2^2 = 0}, \nonumber \\
0 &=& \int\limits_{s_0}^\infty d s  \, \frac{1}{(s + Q_1^2)^2} 
\left[ \sigma_\parallel + \sigma_{LT} + \frac{(s + Q_1^2)}{Q_1 Q_2} \tau^a_{TL} 
\right]_{Q_2^2 = 0}, 
\nonumber \\
0 &=& \int\limits_{s_0}^\infty d s  \, \left[ \frac{\tau_{TL} }{Q_1 Q_2} 
\right]_{Q_2^2 = 0}. \nonumber 
\end{eqnarray}
In these sum rules the $\gamma^\ast \gamma$ fusion cross sections are for one (quasi-) real photon 
and a second virtual photon which can have arbitrary (space-like) virtuality. 
The first  of the sum rules generalizes the GDH sum rule for the helicity-difference $\gamma \gamma$ fusion 
cross section to the case of one real and one virtual photon. The two further sum rules are for 
$\gamma^\ast \gamma$ fusion cross sections which involve longitudinal photon amplitudes. 

We have shown that these sum rules are exactly verified 
for the tree level scalar and spinor QED cross sections. Verifications beyond the tree-level in various 
field theories are underway~\cite{Vlad}. 

We have performed a detailed quantitative study of the new sum rules for the case 
of the production of light quark mesons as well as for the production of mesons in the charm quark sector.  
Using the empirical information in evaluating the sum rules,  
we have found that the helicity-difference sum rule requires cancellations between different 
mesons, implying non-perturbative relations. For the light quark isovector mesons, the $\pi^0$ contribution 
was found to be compensated to around 70\% by the contribution 
of the lowest lying isovector tensor meson $a_2(1320)$. 
For the isoscalar light quark mesons, the $\eta$ and $\eta^\prime$ contributions 
were found to be entirely compensated 
within the experimental accuracy by the two lowest-lying tensor mesons $f_2(1270)$ and $f_2^\prime(1525)$. 
In the charm quark sector, the situation is different as it involves the narrow resonance contributions below 
$D \bar D$ threshold, and the continuum contribution above $D \bar D$ threshold. For the narrow resonances, 
the $\eta_c$ was found to give by far the dominant contribution. When using a duality estimate for the continuum 
contribution, we found that it entirely cancels the narrow resonance contributions, 
verifying the sum rule, and pointing to large tensor strength 
(helicity 2) in the cross sections above $D \bar D$ threshold. It will be interesting to test this property experimentally. 

The helicity difference sum rule has also been applied for the case of one real and one virtual photon. 
In this case the $\gamma^\ast \gamma$ fusion cross sections depend on the meson 
transition form factors (FFs). 
We have reviewed the general formalism and parameterization for the $\gamma^\ast \gamma \to {\rm meson}$ transition FFs for (pseudo-) scalar, axial-vector, and tensor mesons. 
Because for scalar and tensor mesons the $\gamma^\ast \gamma$ transition FFs have not yet been measured, 
a direct test of the sum rules for finite virtuality is not possible at present. 
However, we were able to show that the helicity-difference sum rule allows to provide an estimate for the 
$f_2(1270)$ tensor FF in terms of the $\eta$, and $\eta^\prime$ FFs, and for the $a_2(1320)$ tensor FF in terms of the $\pi^0$ FF. 
Since empirical information on pseudo-scalar meson FFs is available,  
these relations provide predictions for tensor meson FFs which will be interesting to confront with experiment. 

The second new sum rule derived in this paper, involving the $\sigma_\parallel, \sigma_{LT}$, and $\tau^a_{TL}$ $\gamma^\ast \gamma$ response functions, has also been tested 
for the case of quasi-real photons. As pseudo-scalar mesons cannot contribute to this sum rule, 
a cancellation between scalar and tensor mesons on one hand and axial-vector mesons on the other hand 
is at work. 
Using the existing empirical information for quasi-real photons, the contribution of the 
two lowest lying axial-vector mesons $f_1 (1285)$ and $f_1(1420)$ was found to be entirely cancelled, 
within error bars, by the  contribution of the dominant tensor meson $f_2 (1270)$.
When applying this sum rule to the case of one virtual photon, it again allows one to relate 
the $f_2(1270)$ tensor FF, this time to the transition FFs for the $f_1 (1285)$ and $f_1(1420)$ mesons, 
which have both been measured.  
The predictions from the two different sum rules for the $f_2(1270)$ FF were found to agree within 20\% for 
a virtuality below 1 GeV$^2$, and within 35\% up to about 2 GeV$^2$. 

Besides the three super-convergence relations, 
we have also derived sum rules which express the coefficients in a low-energy expansion of the forward light-by-light 
scattering amplitude in terms of $\gamma^\ast \gamma \to X$ cross sections. 
These evaluations may be used as a cross-check for models of the non-forward light-by-light scattering which are applied  
to evaluate the hadronic  LbL contribution to $(g - 2)_\mu$.  

On the experimental side, the ongoing $\gamma \gamma$ physics programs by the  BABAR and Belle Collaborations, 
as well as the upcoming $\gamma \gamma$ physics program by the  BES-III Collaboration, will allow to 
further improve the data situation significantly. In particular, the extraction of the $\gamma^\ast \gamma$ response functions through 
their different azimuthal angular dependencies, and the measurements of multi-meson final states ($\pi \pi$, $\pi \eta, \ldots$) 
promise to access besides the pseudo-scalar meson FFs also the 
scalar, axial-vector and tensor meson FFs, thus allowing direct tests of the sum rule predictions presented in this work.

\section*{Acknowledgements}

This work was supported by the Deutsche Forschungsgemeinschaft DFG through the Collaborative 
Research Center ``The Low-Energy Frontier of the Standard Model" (SFB 1044). 
Furthermore, the work of V. Pauk is also supported by the graduate school Graduate School ``Symmetry Breaking in Fundamental Interactions" 
(DFG/GRK 1581).

\appendix

\section{Kinematics and cross sections  of the $e^\pm + e^- \to e^\pm + e^- + X$ process}
\label{app:gaga}

The kinematics of the process 
$e (p_1) + e (p_2) \to e (p^\prime_1) + e (p^\prime_2) + X$, with $X$ the produced hadronic state, 
in the lepton c.m. system, i.e. the c.m. system of the colliding beams (which we denote by  {\it c.m. ee}) is characterized by the four-vectors of the incoming leptons~:
\begin{eqnarray}
p_1 (E, \vec p_1), \quad \quad \quad  p_2 (E, - \vec p_1), 
\end{eqnarray}
with beam energy $E = \sqrt{s} / 2$, and $s = (p_1 + p_2)^2$. 
\\
The kinematics of the outgoing leptons can be related to the virtual photon four-momenta as~:
\begin{eqnarray}
q_1 = p_1 - p^\prime_1, \quad \quad  \quad  q_2 = p_2 - p^\prime_2 . 
\end{eqnarray}
The kinematics of the outgoing leptons then determines five kinematical quantities~: 
\begin{itemize}
\item
the energies of both virtual photons~:
\begin{eqnarray}
\omega_1 \equiv q_1^0 = E - E^\prime_1, \quad \quad \quad  \omega_2 = q_2^0 \equiv E - E^\prime_2,
\end{eqnarray}
with $E^\prime_1$ and $E^\prime_2$ the energies of both outgoing leptons;

\item
the virtualities of both virtual photons~:
\begin{eqnarray}
Q_1^2 \equiv - q_1^2 = 4 E E^\prime_1 \sin^2 \theta_1 / 2 + Q_{1, \, min}^2 \; ,  
\quad \quad  \quad
Q_2^2 \equiv - q_2^2 = 4 E E^\prime_2 \sin^2 \theta_2 / 2 + Q_{2, \, min}^2 \; , 
\end{eqnarray}
where $\theta_1$ and $\theta_2$ are the (polar) angles of the scattered electrons relative to the 
respective beam directions, and where the minimal values of the virtualities are given by (in the limit where $E^\prime_1 > > m$ and $E^\prime_2 > > m$, with $m$ the lepton mass)~:
\begin{eqnarray}
Q_{1, \, min}^2 \simeq m^2 \frac{\omega_1^2}{E E^\prime_1},
\quad \quad  \quad
Q_{2, \, min}^2 \simeq m^2 \frac{\omega_2^2}{E E^\prime_2};
\end{eqnarray}

\item 
the azimuthal angle $\phi$ between both lepton planes, which in the lepton c.m. frame  
can be obtained as~:
\begin{eqnarray}
\bigl( \cos \phi \bigr)_{c.m. e e} \equiv - \frac{p^\prime_{1 \perp} \cdot p^\prime_{2 \perp}}{\left[ 
(p^\prime_{1 \perp})^2 \; (p^\prime_{2 \perp})^2 \right]^{1/2}},
\end{eqnarray} 
where $p^\prime_{1 \perp}$ and $p^\prime_{2 \perp}$ denote the components of the outgoing lepton four-vectors which are perpendicular to the respective beam directions, and are defined 
in the lepton c.m. frame  as~:
\begin{eqnarray}
\left ( p^\prime_{1 \perp} \right)^\mu = - R^{\mu \nu} (p_1, p_2) \, \left ( p^\prime_1 \right)_\nu , 
\quad \quad  \quad
\left( p^\prime_{2 \perp} \right)^\mu = - R^{\mu \nu} (p_1, p_2) \, \left ( p^\prime_2 \right)_\nu , 
\end{eqnarray}
with
\begin{eqnarray}
R^{\mu \nu} (p_1, p_2) = - g^{\mu \nu} + \frac{1}{\left[ (p_1 \cdot p_2)^2 - m^4 \right]} \;
\bigl \{ (p_1 \cdot p_2) \left( p_1^\mu \, p_2^\nu + p_2^\mu \, p_1^\nu \right)
- m^2 \left( p_1^\mu \, p_1^\nu + p_2^\mu \, p_2^\nu \right)
\bigr \}. 
\end{eqnarray}

\end{itemize}

In the following it will also turn out to be useful to determine kinematical quantities in the 
c.m. system of the virtual photons ( which we denote by {\it c.m. $\gamma \gamma$}). 
In particular, the azimuthal angle between both lepton planes, in the $\gamma \gamma$ c.m. frame,  which we denote by $\tilde \phi$ is given by~:
\begin{eqnarray}
 \cos \tilde \phi  \equiv - \frac{\tilde p_{1 \perp} \cdot \tilde p_{2 \perp}}{\left[ 
(\tilde p_{1 \perp})^2 \; (\tilde p_{2 \perp})^2 \right]^{1/2}},
\label{eq:costilde}
\end{eqnarray} 
where $\tilde p_{1 \perp}$ and $\tilde p_{2 \perp}$ denote the transverse components of the incoming lepton four-vectors in the $\gamma \gamma$  c.m. frame  and are defined in a covariant way as~:
\begin{eqnarray}
\left ( \tilde p_{1 \perp} \right)^\mu = - R^{\mu \nu} (q_1, q_2) \, \left ( p_1 \right)_\nu , 
\quad \quad  \quad
\left(\tilde  p_{2 \perp} \right)^\mu = - R^{\mu \nu} (q_1, q_2) \, \left ( p_2 \right)_\nu , 
\end{eqnarray}
with
\begin{eqnarray}
R^{\mu \nu} (q_1, q_2) = - g^{\mu \nu} + \frac{1}{\left[ (q_1 \cdot q_2)^2 - q_1^2 q_2^2 \right]} \;
\bigl \{ (q_1 \cdot q_2) \left( q_1^\mu \, q_2^\nu + q_2^\mu \, q_1^\nu \right)
- q_1^2 \, q_2^\mu \, q_2^\nu  - q_2^2 \, q_1^\mu \, q_1^\nu 
\bigr \}. 
\end{eqnarray}
As the {\it rhs} of Eq.~(\ref{eq:costilde}) is expressed in a Lorentz invariant way, one can 
then evaluate all four-momenta in the lepton c.m. frame, 
to obtain the expression of $\cos \tilde \phi$ in terms of the lepton c.m. kinematics. 

The cross section for the process 
$e (p_1) + e (p_2) \to e (p^\prime_1) + e (p^\prime_2) + X$, with $X$ the produced hadronic state, 
can be expressed in terms of eight cross sections for the $\gamma^\ast \gamma^\ast \to X$ 
process, which where defined in Eq.~(\ref{eq:vcross}),  as~:  
\begin{eqnarray}
d \sigma &=& \frac{\alpha^2}{16 \pi^4 \, Q_1^2 \, Q_2^2} \, \frac{2 \sqrt{X}}{s (1 - 4 m^2 / s)}  
\cdot \frac{d^3 \vec p_1^{\, \prime}}{E_1^{\prime}} 
\cdot \frac{d^3 \vec p_2^{\, \prime}}{E_2^\prime} \nonumber \\
&\times& \left\{ 
4 \, \rho_1^{++} \, \rho_2^{++} \, \sigma_{TT} 
+  \rho_1^{00} \, \rho_2^{00} \, \sigma_{LL} 
+ 2 \, \rho_1^{++} \, \rho_2^{00} \, \sigma_{TL} 
+ 2 \, \rho_1^{00} \, \rho_2^{++} \, \sigma_{LT} 
\right. \nonumber \\
&&+ 2 \, \left( \rho_1^{++} - 1 \right) \, \left( \rho_2^{++} - 1 \right) \, \left( \cos 2 \tilde \phi \right) \, \tau_{TT} 
+ 8 \, \left[ \frac{\left( \rho_1^{00} + 1 \right) \, \left( \rho_2^{00} + 1 \right)}{\left( \rho_1^{++} - 1 \right) \, \left( \rho_2^{++} - 1 \right)}\right]^{1/2} \, \left( \cos \tilde \phi \right) \, \tau_{TL} 
\nonumber \\
&&\left.
+ h_1 h_2 \, 4  \left[ \left( \rho_1^{00} + 1 \right) \, \left( \rho_2^{00} + 1 \right) \right]^{1/2} \, \tau^a_{TT}
+ h_1 h_2 \, 8  \left[ \left( \rho_1^{++} - 1 \right) \, \left( \rho_2^{++} - 1 \right) \right]^{1/2} \, \left( \cos \tilde \phi \right) \, \tau^a_{TL}
\right\},
\label{eq:gagacross}
\end{eqnarray}
where $h_1 = \pm 1$ and $h_2 = \pm 1$ are both lepton beam helicities, and 
where we have defined kinematical coefficients~:
\begin{eqnarray}
\rho_1^{++} &=& \frac{1}{2} \left\{ 1 - \frac{4 m^2}{Q_1^2} + \frac{1}{X} \left( 2 \, p_1 \cdot q_2 - \nu \right)^2 \right\}\, , \nonumber \\
\rho_2^{++} &=& \frac{1}{2} \left\{ 1 - \frac{4 m^2}{Q_2^2} + \frac{1}{X} \left( 2 \, p_2 \cdot q_1 - \nu \right)^2 \right\}\, , \nonumber \\
\rho_1^{00} &=& \frac{1}{X} \left( 2 \, p_1 \cdot q_2 - \nu \right)^2 - 1 \, , \nonumber \\
\rho_2^{00} &=& \frac{1}{X} \left( 2 \, p_2 \cdot q_1 - \nu \right)^2 - 1\, . 
\label{eq:kincoeff}
\end{eqnarray}

\section{Tree-level $\gamma^\ast \gamma^\ast$ cross sections in QED}
\label{app:perturb}
\subsection{Scalar QED}
\label{app:scalar}

The $\gamma^\ast \gamma^\ast \to {\cal S} \bar {\cal S}$ cross sections (with ${\cal S}$ an electrically charged structureless scalar particle) to lowest order in $\alpha$ are given by~:
\begin{eqnarray}
\sigma_0 + \sigma_2 &=& \sigma_\parallel + \sigma_\perp \nonumber \\
&=& \alpha^2 \frac{\pi}{2} \frac{s^2 \nu^3}{X^3} \left\{ 
\sqrt{a} \left[ 2 - a - \left( 1 - \frac{2 X}{s \nu}  \right)^2 \right]  
 - \left( 1 - a \right)  \left(  3 - \frac{4 X}{s \nu} + a \right) L
\right\}, \\
\sigma_\parallel - \sigma_\perp 
&=& \alpha^2 \frac{\pi}{4} \frac{s^2 \nu^3}{X^3} \left\{ 
\sqrt{a} \left[ 1 - a  + 2 \left( 1 -  \frac{2 X}{s \nu} \right)^2 \right] - \left( 1 - a \right)  
\left( 3 - \frac{8 X}{s \nu} + a \right)  L
\right\}, \\
\sigma_0 - \sigma_2  
&=& \alpha^2 2 \pi \frac{s \nu^2}{X^2} \left\{ 
- \sqrt{a} \left( 1 - \frac{X}{s \nu} \right)  + \left( 1 - a \right)
 L  \right\}, \\
\sigma_{LL}  
&=& \alpha^2 \pi Q_1^2 Q_2^2 \frac{s^2 \nu}{X^3} \left\{ 
\sqrt{a} \left[ 2 + \frac{1}{1 - a} \left( 1 - \frac{X}{s \nu}\right)^2  \right] - \left( 3 +   \frac{X}{s \nu} \right) \left( 1 -   \frac{X}{s \nu} \right)
 L  \right\}, \\
\sigma_{LT}  
&=& \alpha^2 \frac{\pi}{2} Q_1^2 \frac{s \nu (\nu - Q_2^2)^2}{X^3} \left\{ 
- 3 \sqrt{a} + \left( 3 -  a \right)
 L  \right\}, \\
\tau_{TL}  
&=& \alpha^2 \frac{\pi}{2} Q_1 Q_2 \frac{s \nu}{X^2} \left\{ 
- \sqrt{a} + \left( 1 - \frac{2 X}{s \nu} + a \right)
 L  \right\}, \\
\tau^a_{TL}  
&=& \alpha^2 \frac{\pi}{2} Q_1 Q_2 \frac{s^2 \nu^2}{X^3} \left\{ 
\sqrt{a} \left(3 - \frac{4 X}{s \nu} \right) - \left[ 1 - a + 2 \left( 1 - \frac{X}{s \nu} \right)^2 \right]
 L  \right\}, 
\end{eqnarray}
with 
\begin{eqnarray}
L \equiv \ln \left( \frac{1 + \sqrt{a} }{\sqrt{1 - a } } \right), 
\quad \quad \quad 
a \equiv \frac{X}{\nu^2} \left( 1 - \frac{4 m^2}{s} \right). 
\end{eqnarray}
In the limit where one of the virtual photons becomes real ($Q_2^2 = 0$) in case of the response functions involving only transverse photons, 
or becomes quasi-real ($Q_2^2 \approx 0$) in case of the response functions 
involving a longitudinal photon, the above expressions simplify to~:
\begin{eqnarray}
\left[ \sigma_\parallel + \sigma_\perp \right]_{Q_2^2 = 0} &=& \alpha^2 4\pi  \frac{s^2}{(s + Q_1^2)^3} \left\{ 
\sqrt{1 - \frac{4 m^2}{s} } \left( 1 + \frac{4 m^2}{s} + \frac{Q_1^4}{s^2}   \right)  
 - \frac{8 m^2}{s}   \left(  1 -  \frac{2 m^2}{s}   -  \frac{Q_1^2}{s}  \right) L
\right\}, \\
\left[ \sigma_\parallel - \sigma_\perp \right]_{Q_2^2 = 0}
&=& \alpha^2  4 \pi \frac{s^2}{(s + Q_1^2)^3} \left\{ 
\sqrt{1 - \frac{4 m^2}{s} } \left(   \frac{2 m^2}{s}  + \frac{Q_1^4}{s^2} \right)  
+ \frac{8 m^2}{s}  \left(  \frac{m^2}{s} + \frac{Q_1^2}{s} \right)  L
\right\}, \\
\left[ \sigma_0 - \sigma_2  \right]_{Q_2^2 = 0} 
&=& \alpha^2 4 \pi \frac{s}{(s + Q_1^2)^2} \left\{ 
- \sqrt{1 - \frac{4 m^2}{s} } \left( 1 - \frac{Q_1^2}{s} \right)  + \frac{8 m^2}{s} 
 L  \right\}, \\
\left[ \frac{1}{Q_1^2 Q_2^2} \sigma_{LL}  \right]_{Q_2^2 = 0} 
&=& \alpha^2  8 \pi \frac{s^2}{(s + Q_1^2)^5} \left\{ 
 \sqrt{1 - \frac{4 m^2}{s} } \left( 8 + \frac{s}{4 m^2} \left( 1 - \frac{Q_1^2}{s} \right)^2 \right) - \left(7  + \frac{Q_1^2}{s}  \right) 
 \left(1 - \frac{Q_1^2}{s}  \right)
 L  \right\}, \\
\left[ \frac{1}{Q_1^2} \sigma_{LT}  \right]_{Q_2^2 = 0} 
&=& \alpha^2  4 \pi  \frac{s}{(s + Q_1^2)^3} \left\{ 
- 3 \sqrt{1 - \frac{4 m^2}{s} } + 2 \left( 1 + \frac{2 m^2}{s}   \right)
 L  \right\}, \\
\left[ \frac{1}{Q_1 Q_2} \tau_{TL}  \right]_{Q_2^2 = 0} 
&=& \alpha^2  4 \pi \frac{s}{(s + Q_1^2)^3} \left\{ 
- \sqrt{1 - \frac{4 m^2}{s} } + \left( 1 - \frac{Q_1^2}{s} - \frac{4 m^2}{s}   \right)
 L  \right\}, \\
\left[\frac{1}{Q_1 Q_2} \tau^a_{TL}  \right]_{Q_2^2 = 0} 
&=& \alpha^2  8 \pi \frac{s^2}{(s + Q_1^2)^4} \left\{ 
\sqrt{1 - \frac{4 m^2}{s} } \left( 1 - \frac{2 Q_1^2}{s} \right) - \left[ \frac{1}{2} \left(1 - \frac{Q_1^2}{s}\right)^2 + \frac{4 m^2}{s}  \right]
 L  \right\}, 
\end{eqnarray}
 with 
 \begin{eqnarray}
\left[  L \right]_{Q_2^2 = 0}  = \ln \left( \frac{\sqrt{s}}{2 m} \left[ 1 + \sqrt{1 - \frac{4 m^2}{s}} \right] \right), 
\end{eqnarray}

\subsection{Spinor QED}
\label{app:spinor}

The $\gamma^\ast \gamma^\ast \to q \bar q$ cross sections (with $q$ an electrically charged structureless spin-1/2 particle) to lowest order in $\alpha$ are given by~:
\begin{eqnarray}
\sigma_0 + \sigma_2 &=& \sigma_\parallel + \sigma_\perp \nonumber \\
&=& \alpha^2  \pi  \frac{s^2 \nu^3}{X^3} \left\{ 
\sqrt{a} \left[ - 4 \left( 1 - \frac{X}{s \nu} \right)^2 - (1 - a)   
+ \frac{ Q_1^2 Q_2^2}{\nu^2} \left( 2 - \frac{1}{(1 - a)} \, \frac{4 X^2}{s^2 \nu^2} \right) 
\right] \right. \nonumber\\
&&\left. \hspace{1.5cm} + \left[ 3 - a^2 + 2 \left( 1 - \frac{2 X}{s \nu} \right)^2  
- \frac{2 Q_1^2 Q_2^2}{\nu^2} (1 + a) \right] \, L
\right\}, \\
\sigma_\parallel - \sigma_\perp 
&=& \alpha^2 \frac{\pi}{2} \frac{s^2 \nu^3}{X^3} \left\{ 
\sqrt{a} \left[ - (1 - a) - 2 \left( 1 - \frac{2 X}{s \nu} \right)^2  \right] \right. \nonumber \\
&&\left. \hspace{1.5cm} + \left[ - (1 - a)^2 + 4 (1 - a) \left( 1 - \frac{2 X}{s \nu} \right) 
+ \frac{ Q_1^2 Q_2^2}{\nu^2} \, \frac{8 X^2}{s^2 \nu^2}
\right] \, L\right\}, \\
\sigma_0 - \sigma_2  
&=& \alpha^2 4 \pi \frac{s \nu^2}{X^2} \left\{ 
\sqrt{a} \left[ 2 - \frac{X}{s \nu}    
- \frac{ Q_1^2 Q_2^2}{\nu^2}  \frac{1}{(1 - a)} \, \frac{X}{s \nu}  \right]
\, - 2   \left( 1 - \frac{X}{s \nu} \right) \, L 
 \right\}, \\
\sigma_{LL}  
&=& \alpha^2 2 \pi Q_1^2 Q_2^2 \frac{s^2}{\nu X^2} \left\{ 
\sqrt{a} \left[ -2  -  \frac{(3 - 2 a)}{(1 - a)} \frac{Q_1^2 Q_2^2}{X}\right]
+ \left( 2 + \frac{3 Q_1^2 Q_2^2}{X} 
\right) \, L
  \right\}, \\
\sigma_{LT}  
&=& \alpha^2 \pi Q_1^2 \frac{s}{\nu X^2} \left\{
\sqrt{a} \left[ (\nu - Q_2^2)^2 \left( 2 + \frac{3 Q_1^2 Q_2^2}{X}\right)
- 2 \nu Q_2^2 + Q_2^4 \frac{(3 - a)}{(1 - a)}
\right]  \right. \nonumber \\
&&\left.\hspace{1.5cm}+ 
\left[ (\nu - Q_2^2)^2 \left( -2 (1 - a) - (3 - a) \frac{Q_1^2 Q_2^2}{X} \right) 
+ 2 \nu Q_2^2 (1 + a) - Q_2^4 (3 + a)
\right] \, L
  \right\}, \\
\tau_{TL}  
&=& \alpha^2  2 \pi  \left( Q_1 Q_2 \right)^{3} \frac{s}{\nu X^2} \left\{ 
\frac{\sqrt{a}}{1 - a} - L  \right\}, \\
\tau^a_{TL}  
&=& \alpha^2  \pi  Q_1 Q_2 \frac{s^2 \nu^2}{X^3} \left\{ 
- \sqrt{a} \left( 3 - \frac{4 X}{s \nu} \right) + \left(3 - \frac{4 X}{s \nu}  - a  \right) L  \right\}, 
\end{eqnarray}
with 
\begin{eqnarray}
L \equiv \ln \left( \frac{1 + \sqrt{a} }{\sqrt{1 - a } } \right), 
\quad \quad \quad 
a \equiv \frac{X}{\nu^2} \left( 1 - \frac{4 m^2}{s} \right). 
\end{eqnarray}
In the limit where one of the virtual photons becomes real ($Q_2^2 = 0$) in case of the response functions involving only transverse photons, 
or becomes quasi-real ($Q_2^2 \approx 0$) in case of the response functions 
involving a longitudinal photon, the above expressions simplify to~:
\begin{eqnarray}
\left[ \sigma_\parallel + \sigma_\perp \right]_{Q_2^2 = 0} &=& \alpha^2 8 \pi  \frac{s^2}{(s + Q_1^2)^3} 
\left\{ 
\sqrt{1 - \frac{4 m^2}{s} } \left[ - \left( 1 - \frac{Q_1^2}{s} \right)^2 - \frac{4 m^2}{s} \right] 
+ 2 \left( 1 + \frac{4 m^2}{s} - \frac{8 m^4}{s^2}   + \frac{Q_1^4}{s^2} \right) L 
\right\}, \\
\left[ \sigma_\parallel - \sigma_\perp \right]_{Q_2^2 = 0}
&=& - \, \alpha^2  8 \pi \frac{s^2}{(s + Q_1^2)^3} \left\{ 
\sqrt{1 - \frac{4 m^2}{s} } \left(  \frac{2 m^2}{s} + \frac{Q_1^4}{s^2} \right)
+ \frac{8 m^2}{s} \left( \frac{m^2}{s} +  \frac{Q_1^2}{s} \right)  L  
\right\}, \\
\left[ \sigma_0 - \sigma_2  \right]_{Q_2^2 = 0} 
&=& \alpha^2 8 \pi \frac{s}{(s + Q_1^2)^2} \left\{ 
\sqrt{1 - \frac{4 m^2}{s} } \left( 3 - \frac{Q_1^2}{s} \right) - 2 \left( 1 - \frac{Q_1^2}{s}  \right)
 L   \right\}, 
 \label{eq:spinheldiff}
 \\
\left[ \frac{1}{Q_1^2 Q_2^2} \sigma_{LL}  \right]_{Q_2^2 = 0} 
&=& \alpha^2  128 \pi \frac{s^2}{(s + Q_1^2)^5} \left\{ 
- \sqrt{1 - \frac{4 m^2}{s} } + L
   \right\}, \\
\left[ \frac{1}{Q_1^2} \sigma_{LT}  \right]_{Q_2^2 = 0} 
&=& \alpha^2  16 \pi  \frac{s}{(s + Q_1^2)^3} \left\{
 \sqrt{1 - \frac{4 m^2}{s} } - \frac{4 m^2}{s} \, L
 \right\}, \\
\left[ \frac{1}{Q_1 Q_2} \tau_{TL}  \right]_{Q_2^2 = 0} 
&=& 0, \\
\left[\frac{1}{Q_1 Q_2} \tau^a_{TL}  \right]_{Q_2^2 = 0} 
&=& \alpha^2  16 \pi \frac{s^2}{(s + Q_1^2)^4} \left\{ 
- \sqrt{1 - \frac{4 m^2}{s} } \left( 1 - \frac{2 Q_1^2}{s} \right) + \left( - \frac{2 Q_1^2}{s} + \frac{4 m^2}{s}  \right)
 L  \right\}, 
\end{eqnarray}
 with 
 \begin{eqnarray}
\left[  L \right]_{Q_2^2 = 0}  = \ln \left( \frac{\sqrt{s}}{2 m} \left[ 1 + \sqrt{1 - \frac{4 m^2}{s}} \right] \right).
\end{eqnarray}

\section{$\gamma^\ast \gamma^\ast \to$~meson transition form factors}
\label{sec7}

In this Appendix we detail the formalism and the available data for the
$\gamma^\ast \gamma^\ast \to$~meson transition form factors (FFs), and 
 successively discuss the $C$-even
pseudo-scalar ($J^{PC} = 0^{-+}$), scalar ($J^{PC} = 0^{++}$), axial-vector ($J^{PC} = 1^{++}$), 
and tensor ($J^{PC} = 2^{++}$) mesons.

\subsection{Pseudo-scalar mesons}

The process 
$\gamma^\ast (q_1, \lambda_1) + \gamma^\ast(q_2, \lambda_2) \to {\cal P}$, 
describing the transition from an initial state of two  
virtual photons, with four-momenta $q_1, q_2$ and helicities $\lambda_1, \lambda_2 = 0, \pm 1$, 
to a pseudo-scalar meson ${\cal P} = \pi^0, \eta, \eta^\prime, \eta_c, ...$ ($J^{PC} = 0^{-+}$) with mass $m_P$, is  described by the matrix element~:
\begin{eqnarray} 
{\cal M}(\lambda_1, \lambda_2) = - i \, e^2 \, \varepsilon_{\mu \nu \alpha \beta} \, 
\varepsilon^\mu(q_1, \lambda_1) \, \varepsilon^\nu(q_2, \lambda_2) \, 
q_1^\alpha \, q_2^{\beta} \, 
F_{{\cal P} \gamma^\ast \gamma^\ast} (Q_1^2, Q_2^2), 
\label{eq:psff}
\end{eqnarray}
where $ \varepsilon^\alpha(q_1, \lambda_1)$ and $\varepsilon^\beta(q_2, \lambda_2)$ 
are the polarization vectors of the virtual photons, and where the 
meson structure information is encoded in the form factor (FF)  $F_{{\cal P} \gamma^\ast \gamma^\ast}$, 
which is a function of the virtualities of both photons, satisfying 
$F_{{\cal P} \gamma^\ast \gamma^\ast} (Q_1^2, Q_2^2) = F_{{\cal P} \gamma^\ast \gamma^\ast} (Q_2^2, Q_1^2)$.   
From Eq.~(\ref{eq:psff}), one can easily deduce that the only non-zero $\gamma^\ast \gamma^\ast \to {\cal P}$ 
helicity amplitudes, which we define in the rest frame of the produced meson, are given by~:
\begin{eqnarray}
{\cal M}(\lambda_1 = +1, \lambda_2 = +1) = - {\cal M}(\lambda_1 = -1, \lambda_2 = -1) = 
- e^2 \, \sqrt{X} \, F_{{\cal P} \gamma^\ast \gamma^\ast}(Q_1^2, Q_2^2) \, . 
\end{eqnarray}
The FF at $Q_1^2 = Q_2^2 = 0$, $F_{{\cal P} \gamma^\ast \gamma^\ast}(0, 0)$, describes the  
two-photon decay width of the pseudo-scalar meson~:
\begin{eqnarray}
\Gamma_{\gamma \gamma}({\cal P}) = \frac{\pi \alpha^2}{4} \, m_P^3 \, 
| F_{{\cal P} \gamma^\ast \gamma^\ast}(0, 0)  | ^2,
\label{ps2gwidth}
\end{eqnarray}
with $m_P$ the pseudo-scalar meson mass, and $\alpha = e^2 / (4 \pi) \simeq 1/137$.

In this paper, we study the sum rules involving cross sections for one real photon and one virtual photon.  
For one real photon ($Q_2^2 = 0$), the only non-vanishing cross sections in 
Eq.~(\ref{eq:vcross}) are given by~:
\begin{eqnarray}
\left[ \sigma_0 \right]_{Q_2^2 = 0}  = \left[ \sigma_\perp \right]_{Q_2^2 = 0}  
= 2 \left[ \sigma_{TT} \right]_{Q_2^2 = 0}  = - \left[ \tau_{TT} \right]_{Q_2^2 = 0}  
= \delta(s - m_P^2) \, 16 \, \pi^2 \, \frac{\Gamma_{\gamma \gamma} ( {\cal P})}{m_P} \, 
\left(1 + \frac{Q_1^2}{m_P^2} \right) \, \left[ \frac{F_{{\cal P} \gamma^\ast \gamma^\ast}(Q_1^2, 0)}{F_{{\cal P} \gamma^\ast \gamma^\ast}(0, 0)} \right]^2\, .
\label{eq:pscross}
\end{eqnarray}

For massless quarks, the divergence of the isovector axial current,  
$A_3^\mu \equiv \frac{1}{\sqrt{2}}(\bar u \gamma^\mu \gamma_5 u - \bar d \gamma^\mu \gamma_5  d)$, 
does not vanish but 
exhibits an anomaly due to the triangle graphs which allow the 
$\pi^0$ to couple to two vectors currents (Wess-Zumino-Witten anomaly).  
For the $\pi^0$, the chiral (isovector axial) anomaly, 
predicts that its transition FF at $Q_1^2 = Q_2^2 = 0$ is given by~:
\begin{eqnarray}
F_{\pi^0 \gamma^\ast \gamma^\ast}(0, 0) = \frac{1}{4 \pi^2 f_\pi},
\label{chianom}
\end{eqnarray}
where the pion decay constant $f_\pi$ is defined through the 
isovector axial current matrix element~:
\begin{eqnarray} 
\langle 0 |  A_3^\mu(0) | \pi^0 (p) \rangle = i \, ( \sqrt{2} \, f_\pi ) \, p^\mu.
\end{eqnarray}
When using the current empirical value of the pion decay constant $f_\pi \simeq 92.4$~MeV 
to evaluate the chiral anomaly prediction of Eq.~(\ref{chianom}), one 
obtains the value $F_{M \gamma^\ast \gamma}(0) \simeq 0.274$~GeV$^{-1}$, which 
yields through Eq.~(\ref{ps2gwidth}) a $2 \gamma$ decay width in very 
good agreement with the experimental value (see Table~\ref{table_isovector}). 

The form factors $F_{{\cal P} \gamma^\ast \gamma^\ast}(Q_1^2, 0)$ for one virtual photon and one real photon have been measured 
for $\pi^0$, $\eta$, $\eta^\prime$ by the CELLO~\cite{Behrend:1990sr} , CLEO~\cite{Gronberg:1997fj}, 
and BABAR~\cite{Aubert:2009mc,  BABAR:2011ad}  Collaborations, and for $\eta_c(1S)$ by the 
BABAR Collaboration~\cite{Lees:2010de}.  
In the $Q_1^2$ range up to 10 GeV$^2$, a good parameterization of the data is obtained by the monopole form~:
\begin{eqnarray}
\frac{F_{{\cal P} \gamma^\ast \gamma^\ast}(Q_1^2, 0)}{F_{{\cal P} \gamma^\ast \gamma^\ast}(0, 0)} = \frac{1}{1 + Q_1^2 / \Lambda_P^2},
\label{eq:psffmono}
\end{eqnarray}
where $\Lambda_P$ is the monopole mass parameter. In Table~\ref{tab_ps}, we show the experimental extraction of $\Lambda_P$ 
for the $\pi^0, \eta, \eta^\prime$, and $\eta_c(1S)$ mesons.  

\begin{table}[h]
{\centering \begin{tabular}{|c|c|}
\hline
&  $\Lambda_P$   \\
& [MeV] \\
\hline 
\quad $\pi^0$  & \quad  $776 \pm 22 $  \quad   \\
\quad $\eta$  & \quad  $774 \pm 29 $  \quad   \\
\quad $\eta^\prime$  & \quad  $859 \pm 28 $  \quad   \\
\quad $\eta_c(1S)$  & \quad  $2920 \pm 160$  \quad   \\
\hline
\end{tabular}\par}
\caption{Experimental extraction of the monopole mass parameter in the $\gamma^\ast \gamma \to {\cal P}$ form factors, 
according to the fit of Eq.~(\ref{eq:psffmono}). 
The measured value of $\Lambda_P$ for ${\cal P} =  \pi^0, \eta, \eta^\prime$ is from the CLEO Collaboration~\cite{Gronberg:1997fj}. 
For the $\eta_c(1S)$ state, the measured value is from the BABAR Collaboration~\cite{Lees:2010de}.}
\label{tab_ps}
\end{table}

\subsection{Scalar mesons}

We next consider the process 
$\gamma^\ast (q_1, \lambda_1) + \gamma^\ast(q_2, \lambda_2) \to {\cal S}$, 
describing the transition from an initial state of two  
virtual photons, with four-momenta $q_1, q_2$ and helicities $\lambda_1, \lambda_2 = 0, \pm 1$, 
to a scalar meson ${\cal S}$ ($J^{PC} = 0^{++}$) with mass $m_S$. Scalar mesons can be produced either by two transverse photons or by two longitudinal photons~\cite{Poppe:1986dq, Schuler:1997yw}. 
Therefore, the $\gamma^\ast \gamma^\ast \to {\cal S}$ transition can be described by the matrix element~:
\begin{eqnarray} 
{\cal M}(\lambda_1, \lambda_2) &=&  e^2 \, \varepsilon^\mu(q_1, \lambda_1) \, \varepsilon^\nu(q_2, \lambda_2) \, 
\, \nonumber \\
&\times& \left( \frac{\nu}{m_S} \right) \left\{ 
 - R^{\mu \nu} (q_1, q_2) F^T_{{\cal S} \gamma^\ast \gamma^\ast} (Q_1^2, Q_2^2)  \,+\,
\frac{\nu}{X} \left( q_1^\mu + \frac{Q_1^2}{\nu} q_2^{\mu} \right) \left( q_2^\nu + \frac{Q_2^2}{\nu} q_1^{\nu} \right)  
F^L_{{\cal S} \gamma^\ast \gamma^\ast} (Q_1^2, Q_2^2) 
\right\},
\label{eq:sff}
\end{eqnarray}
where we introduced the symmetric transverse tensor $R^{\mu \nu}$~:
\begin{eqnarray}
R^{\mu \nu} (q_1, q_2) \equiv - g^{\mu \nu} + \frac{1}{X} \,
\bigl \{ \nu \left( q_1^\mu \, q_2^\nu + q_2^\mu \, q_1^\nu \right)
+ Q_1^2 \, q_2^\mu \, q_2^\nu  + Q_2^2 \, q_1^\mu \, q_1^\nu 
\bigr \} ,
\end{eqnarray}
which projects onto both transverse photons, having the properties~:
\begin{eqnarray}
q_{1 \mu} R^{\mu \nu} (q_1, q_2) = 0, \quad q_{1 \nu} R^{\mu \nu} (q_1, q_2) = 0, 
\quad q_{2 \mu} R^{\mu \nu} (q_1, q_2) = 0, \quad q_{2 \nu} R^{\mu \nu} (q_1, q_2) = 0.
\nonumber
\end{eqnarray} 
In Eq.~(\ref{eq:sff}), 
the scalar meson structure information is encoded in the form factors  $F^T_{{\cal S} \gamma^\ast \gamma^\ast}$ 
and $F^L_{{\cal S} \gamma^\ast \gamma^\ast}$, 
which are a function of the virtualities of both photons, where the superscripts indicate the situation where either both 
photons are transverse ($T$) or longitudinal ($L$). Note that the pre-factor $\nu/m_S$ in Eq.~(\ref{eq:sff}) is chosen such 
that the FFs are dimensionless.  
Furthermore, both form factors are symmetric under interchange of both virtualities~: 
\begin{eqnarray}
F^{T, L}_{{\cal S} \gamma^\ast \gamma^\ast}(Q_1^2, Q_2^2) &=& F^{T, L}_{{\cal S} \gamma^\ast \gamma^\ast}(Q_2^2, Q_1^2) .
\end{eqnarray}
From Eq.~(\ref{eq:sff}), one can easily deduce that the only non-zero $\gamma^\ast \gamma^\ast \to {\cal S}$ 
helicity amplitudes are given by~:
\begin{eqnarray}
{\cal M}(\lambda_1 = +1, \lambda_2 = +1) &=& {\cal M}(\lambda_1 = -1, \lambda_2 = -1) = 
e^2 \, \frac{\nu}{m_S} \, F^T_{{\cal S} \gamma^\ast \gamma^\ast}(Q_1^2, Q_2^2) \, , \nonumber \\
{\cal M}(\lambda_1 = 0, \lambda_2 = 0) &=& - \, e^2 \, \frac{Q_1 Q_2}{m_S} \, 
F^L_{{\cal S} \gamma^\ast \gamma^\ast}(Q_1^2, Q_2^2) \, . 
\end{eqnarray}
The transverse FF at $Q_1^2 = Q_2^2 = 0$, $F^T_{{\cal S} \gamma^\ast \gamma^\ast}(0,0)$, describes the  
two-photon decay width of the scalar meson~:
\begin{eqnarray}
\Gamma_{\gamma \gamma}({\cal S}) =  \frac{\pi \alpha^2}{4} m_S \, 
| F^T_{{\cal S} \gamma^\ast \gamma^\ast}(0,0)  | ^2.
\label{s2gwidth}
\end{eqnarray}

In this paper, we study the sum rules involving cross sections for one real photon and one virtual photon.  
For one real photon ($Q_2^2 = 0$), the only non-vanishing cross sections in 
Eq.~(\ref{eq:vcross}) are given by~:
\begin{eqnarray}
\left[ \sigma_0 \right]_{Q_2^2 = 0}  = \left[ \sigma_\parallel \right]_{Q_2^2 = 0}  
= 2 \left[ \sigma_{TT} \right]_{Q_2^2 = 0}  = \left[ \tau_{TT} \right]_{Q_2^2 = 0}  
= \delta(s - m_S^2) \, 16 \, \pi^2 \, \frac{\Gamma_{\gamma \gamma}({\cal S})}{m_S} \, 
\left( 1 + \frac{Q_1^2}{m_S^2} \right) 
\, \left[ \frac{F^T_{{\cal S} \gamma^\ast \gamma^\ast}(Q_1^2, 0)}{F^T_{{\cal S} \gamma^\ast \gamma^\ast}(0, 0)} \right]^2 \, .
\label{eq:scross}
\end{eqnarray}

\subsection{Axial-vector mesons}

We next discuss the two-photon production of an axial vector meson. 
Due to the symmetry under rotational invariance, spatial inversion as well as the Bose symmetry of a state of two real photons, 
the production of a spin-1 resonance by two real photons is forbidden, which is known as the Landau-Yang theorem~\cite{Yang:1950rg}. 
However the production of an axial-vector meson by two photons is possible when one or both photons are virtual. 
The matrix element for the process $\gamma^\ast (q_1, \lambda_1) + \gamma^\ast(q_2, \lambda_2) \to {\cal A}$, 
describing the transition from an initial state of two  
virtual photons, with four-momenta $q_1, q_2$ and helicities $\lambda_1, \lambda_2 = 0, \pm 1$, 
to an axial-vector meson ${\cal A}$ ($J^{PC} = 1^{++}$) with mass $m_A$ and helicity $\Lambda = \pm 1, 0$ (defined along the 
direction of $\vec q_1$), 
 is  described by three structures~\cite{Poppe:1986dq, Schuler:1997yw}, and can be parameterized as~:
\begin{eqnarray} 
{\cal M}(\lambda_1, \lambda_2; \Lambda) &=& e^2 \, 
\varepsilon_\mu(q_1, \lambda_1) \, \varepsilon_\nu(q_2, \lambda_2) \, 
\varepsilon^{\alpha \ast}(p_f, \Lambda) \, \nonumber \\
&\times& i \, \varepsilon_{\rho \sigma \tau \alpha} \,  \left\{ 
R^{\mu \rho} (q_1, q_2) R^{\nu \sigma} (q_1, q_2) \, 
(q_1 - q_2)^\tau \, \frac{\nu}{m_A^2} \, F^{(0)}_{{\cal A} \gamma^\ast \gamma^\ast}(Q_1^2, Q_2^2)
\right. \nonumber \\
&&\hspace{1cm} + \, R^{\nu \rho}(q_1, q_2) \left( q_1^\mu + \frac{Q_1^2}{\nu} q_2^{\mu} \right) 
 q_1^\sigma \, q_2^\tau \,  \frac{1}{m_A^2} \, F_{{\cal A} \gamma^\ast \gamma^\ast}^{(1)}(Q_1^2, Q_2^2) \nonumber \\
&&\left. \hspace{1cm} + \, R^{\mu \rho}(q_1, q_2) \left( q_2^\nu + \frac{Q_2^2}{\nu} q_1^{\nu} \right) 
 q_2^\sigma \, q_1^\tau \, \frac{1}{m_A^2} \, F^{(1)}_{{\cal A} \gamma^\ast \gamma^\ast}(Q_2^2, Q_1^2) 
\right\}.
\label{eq:aff}
\end{eqnarray}
In Eq.~(\ref{eq:aff}), 
the axial-vector meson structure information is encoded in the form factors  $F^{(0)}_{{\cal A} \gamma^\ast \gamma^\ast}$ 
and $F^{(1)}_{{\cal A} \gamma^\ast \gamma^\ast}$, where the superscript indicates the helicity state of the axial-vector meson. 
Note that 
only transverse photons give a non-zero transition to a state of helicity zero.
The form factors are functions of the virtualities of both photons, and $F^{(0)}_{{\cal A} \gamma^\ast \gamma^\ast}$ 
is symmetric under the interchange $Q_1^2 \leftrightarrow Q_2^2$. 
In contrast, $F^{(1)}_{{\cal A} \gamma^\ast \gamma^\ast}$ does not need to be symmetric under interchange of 
both virtualities, as can be seen from Eq.~(\ref{eq:aff}). 
 
From Eq.~(\ref{eq:aff}), one can easily deduce that the only non-zero $\gamma^\ast \gamma^\ast \to {\cal A}$ 
helicity amplitudes are given by~:
\begin{eqnarray}
{\cal M}(\lambda_1 = +1, \lambda_2 = +1; \Lambda = 0) &=& - {\cal M}(\lambda_1 = -1, \lambda_2 = -1; \Lambda = 0) = 
e^2 \, (Q_1^2 - Q_2^2) \, \frac{\nu}{m_A^3} \, F^{(0,T)}_{{\cal A} \gamma^\ast \gamma^\ast}(Q_1^2, Q_2^2) \, , \nonumber \\
{\cal M}(\lambda_1 = 0, \lambda_2 = +1; \Lambda = -1) &=& - \, e^2 \, Q_1 \, \left( \frac{X}{\nu m_A^2} \right) \, 
F^{(1)}_{{\cal A} \gamma^\ast \gamma^\ast}(Q_1^2, Q_2^2) \, , \nonumber \\
{\cal M}(\lambda_1 = -1, \lambda_2 = 0; \Lambda = -1) &=& - \, e^2 \, Q_2 \, \left( \frac{X}{\nu m_A^2} \right) \, 
F^{(1)}_{{\cal A} \gamma^\ast \gamma^\ast}(Q_2^2, Q_1^2) \, . 
\end{eqnarray}
Note that the helicity amplitude with two transverse photons vanishes when both photons are real, in accordance with the 
Landau-Yang theorem. 

The matrix element $F^{(1)}_{{\cal A} \gamma^\ast \gamma}(0, 0)$ allows to define an equivalent two-photon decay width 
for an axial-vector meson to decay in one quasi-real longitudinal photon and a (transverse) real photon as~
\footnote{In defining the equivalent two-photon decay width for an axial-vector meson, we follow the convention of 
Ref.~\cite{Schuler:1997yw}, which is also followed in experimental analyses~\cite{Achard:2001uu, Achard:2007hm}. Note 
however that the definition for $\tilde \Gamma_{\gamma \gamma}$ adopted here is one half of that used in Ref.~\cite{Cahn:1986qg}. }:  
\begin{eqnarray}
\tilde \Gamma_{ \gamma \gamma}({\cal A}) \equiv \lim \limits_{Q_1^2 \to 0} \, \frac{m_A^2}{Q_1^2} \, \frac{1}{2} \,
\Gamma \left( {\cal A} \to \gamma^\ast_L \gamma_T \right)
= \frac{\pi \alpha^2}{4} \, m_A \, \frac{1}{3} \left[ F^{(1)}_{{\cal A} \gamma^\ast \gamma^\ast}(0, 0)  \right]^2,
\label{a2gwidth}
\end{eqnarray}
where we have introduced the decay width $\Gamma \left( {\cal A} \to \gamma^\ast_L \gamma_T \right)$ 
for an axial-vector meson to decay in a virtual longitudinal photon, with virtuality $Q_1^2$, 
and a real transverse photon ($Q_2^2 = 0$), as~:
\begin{eqnarray}
\Gamma \left( {\cal A} \to \gamma^\ast_L \gamma_T \right)
= \frac{\pi \alpha^2}{2} \, m_A \, \frac{1}{3} \, \frac{Q_1^2}{m_A^2} \, \, \left(1 + \frac{Q_1^2}{m_A^2} \right)^3 \, 
\left[ F^{(1)}_{{\cal A} \gamma^\ast \gamma^\ast}(Q_1^2, 0)  \right]^2.
\label{a2gwidthlt}
\end{eqnarray}

In this paper, we study the sum rules involving cross sections for one real photon and one virtual photon.  
For one quasi-real photon ($Q_2^2 \to 0$), we can obtain from the above helicity amplitudes and using Eq.~(\ref{eq:abs}) 
the axial-vector meson contributions to the response functions of Eq.~(\ref{eq:vcross}) as~:
\begin{eqnarray}
\left[ \sigma_0 \right]_{Q_2^2 = 0} = \left[ \sigma_\perp \right]_{Q_2^2 = 0}  
&=& 2 \left[ \sigma_{TT} \right]_{Q_2^2 = 0} = - \left[ \tau_{TT} \right]_{Q_2^2 = 0}  
= \delta(s - m_A^2) \, 4 \, \pi^3 \alpha^2 \, \frac{Q_1^4}{m_A^4} \left(1 + \frac{Q_1^2}{m_A^2} \right) \, 
\left[ F^{(0)}_{{\cal A} \gamma^\ast \gamma^\ast }(Q_1^2, 0) \right]^2\, , 
\nonumber \\
\left[ \sigma_{LT} \right]_{Q_2^2 = 0}  
&=&  \delta(s - m_A^2) \, 16 \, \pi^2 \, \frac{3 \, \tilde \Gamma_{\gamma \gamma}({\cal A})}{m_A} \, \frac{Q_1^2}{m_A^2} \, 
\left(1 + \frac{Q_1^2}{m_A^2} \right) \, 
\left[ \frac{F^{(1)}_{{\cal A} \gamma^\ast \gamma^\ast }(Q_1^2, 0)}{F^{(1)}_{{\cal A} \gamma^\ast \gamma^\ast}(0, 0) } \right]^2\, , 
\nonumber \\
\left[ \tau_{TL} \right]_{Q_2^2 = 0}  = - \left[ \tau^a_{TL} \right]_{Q_2^2 = 0}  
&=& \delta(s - m_A^2) \, 8 \, \pi^2  
\, \frac{3 \, \tilde \Gamma_{\gamma \gamma}({\cal A})}{m_A} \, \frac{Q_1 Q_2}{m_A^2} \, \left(1 + \frac{Q_1^2}{m_A^2} \right) \, 
\left[ \frac{F^{(1)}_{{\cal A} \gamma^\ast \gamma^\ast }(Q_1^2, 0)}{F^{(1)}_{{\cal A} \gamma^\ast \gamma^\ast}(0, 0)} 
\cdot 
\frac{F^{(1)}_{{\cal A} \gamma^\ast \gamma^\ast }(0, Q_1^2)}{F^{(1)}_{{\cal A} \gamma^\ast \gamma^\ast}(0, 0)}   \right] \, . 
\label{eq:across}
\end{eqnarray}
Extracting the FFs $F^{(1)}$, and $F^{(0)}$ separately from experiment requires the measurements of $\sigma_{LT}$ and $\sigma_{TT}$ respectively. As experiments to date have not achieved this separation, one is so far only sensitive to the quantity 
$\sigma_{TT} + \varepsilon_1 \, \sigma_{LT}$, where $\varepsilon_1$ is a kinematical parameter (so-called virtual photon polarization parameter) defined as $\varepsilon_1 \equiv \rho_1^{00} / 2 \rho_1^{++}$, see Appendix~\ref{app:gaga}. 
Note that in high-energy collider experiments, 
one typically has $\varepsilon_1 \approx 1$. 
From Eq.~(\ref{eq:across}) one then obtains for this experimentally accessible combination~:
 \begin{eqnarray}
\left[  \sigma_{LT} \left( 1 + \frac{1}{\varepsilon_1} \frac{\sigma_{TT}}{\sigma_{LT}}\right) \right]_{Q_2^2 = 0}   &=& 
  \delta(s - m_A^2) \, 16 \, \pi^2 \, \frac{3 \, \tilde \Gamma_{\gamma \gamma}({\cal A})}{m_A} \, \frac{Q_1^2}{m_A^2} \, 
\left(1 + \frac{Q_1^2}{m_A^2} \right) \, \nonumber \\
&\times& \left( 
\left[ \frac{F^{(1)}_{{\cal A} \gamma^\ast \gamma^\ast }(Q_1^2, 0)}{F^{(1)}_{{\cal A} \gamma^\ast \gamma^\ast}(0, 0) } \right]^2\,  
+ \frac{1}{\varepsilon_1} \, \frac{Q_1^2}{2 \, m_A^2} \, \left[ \frac{F^{(0)}_{{\cal A} \gamma^\ast \gamma^\ast }(Q_1^2, 0)}{F^{(1)}_{{\cal A} \gamma^\ast \gamma^\ast}(0, 0) } \right]^2\,  
\right), 
 \label{eq:expaxff} 
 \end{eqnarray}

We can compare the above general formalism for the two-photon production of an axial-vector meson with the 
description of Ref.~\cite{Cahn:1986qg}, which is commonly used in the literature, and is 
based on a non-relativistic quark model calculation leading to 
only one independent amplitude  for the $\gamma^\ast  \gamma^\ast \to {\cal A}$ process as~:
\begin{eqnarray} 
{\cal M}(\lambda_1, \lambda_2; \Lambda) &=&  e^2 \, 
\varepsilon^\mu(q_1, \lambda_1) \, \varepsilon^\nu(q_2, \lambda_2) \, 
\varepsilon^{\alpha \ast}(p_f, \Lambda) \, 
\, i \varepsilon_{\mu \nu \tau \alpha} \,  
\left( -Q_1^2 \, q_2^\tau + Q_2^2 \, q_1^\tau \right) \, A(Q_1^2, Q_2^2),
\label{eq:affcahn}
\end{eqnarray}
where the independent form factor $A$ satisfies~: $A(Q_1^2, Q_2^2) = A(Q_2^2, Q_1^2)$. 
In such a non-relativistic quark model limit, we can recover Eq.~(\ref{eq:affcahn}) from Eq.~(\ref{eq:aff}) through the 
identifications~:
\begin{eqnarray}
F^{(0)}(Q_1^2, Q_2^2) &=& m_A^2 \, A(Q_1^2, Q_2^2), \nonumber \\
F^{(1)}(Q_1^2, Q_2^2) &=& - \frac{\nu}{X} (\nu + Q_2^2) \, m_A^2 \, A(Q_1^2, Q_2^2), \nonumber \\
F^{(1)}(Q_2^2, Q_1^2) &=& - \frac{\nu}{X} (\nu + Q_1^2) \, m_A^2 \, A(Q_1^2, Q_2^2),
\label{eq:axffcahn}
\end{eqnarray}
in which $2 \nu = m_A^2 + Q_1^2 + Q_2^2$. 
In such model, the experimentally measured two-photon cross section combination of Eq.~(\ref{eq:expaxff}), where $Q_2^2 = 0$, 
is proportional to~:
\begin{eqnarray}
\left[  \sigma_{LT} \left( 1 + \frac{1}{\varepsilon_1} \frac{\sigma_{TT}}{\sigma_{LT}}\right) \right]_{Q_2^2 = 0}   = 
  \delta(s - m_A^2) \, 16 \, \pi^2 \, \frac{3 \, \tilde \Gamma_{\gamma \gamma}({\cal A})}{m_A} \, \frac{Q_1^2}{m_A^2} \, 
\left(1 + \frac{Q_1^2}{m_A^2} \right) \, 
\left( 1 + \frac{1}{\varepsilon_1} \, \frac{Q_1^2}{2 \, m_A^2} \right) \, 
\left[ \frac{A(Q_1^2, 0)}{A(0, 0) } \right]^2.   
 \label{eq:expaxffcahn} 
 \end{eqnarray}
To apply this formula to experimental results where the axial-vector meson has a finite width, one commonly replaces the delta-function 
in Eq.~(\ref{eq:expaxffcahn}) by a Breit-Wigner form, yielding~:
\begin{eqnarray}
\left[  \sigma_{LT} \left( 1 + \frac{1}{\varepsilon_1} \frac{\sigma_{TT}}{\sigma_{LT}}\right) \right]_{Q_2^2 = 0}   = 
48 \, \pi \, \frac{\tilde \Gamma_{\gamma \gamma}({\cal A}) \, \Gamma_{total}}{(s - m_A^2)^2 + m_A^2 \, \Gamma^2_{total}}   \, 
\frac{Q_1^2}{m_A^2} \, 
\left(1 + \frac{Q_1^2}{m_A^2} \right) \, 
\left( 1 + \frac{1}{\varepsilon_1} \, \frac{Q_1^2}{2 \, m_A^2} \right) \, 
\left[ \frac{A(Q_1^2, 0)}{A(0, 0) } \right]^2,  
 \label{eq:expaxffcahn2} 
 \end{eqnarray}
where $\Gamma_{total}$ is the total decay width of the axial-vector meson. 

Phenomenologically, the two-photon production cross sections have been measured for the two lowest lying axial-vector mesons~: 
$f_1(1285)$ and $f_1(1420)$. The most recent measurements were performed by the L3 Collaboration~\cite{Achard:2001uu, Achard:2007hm}. 
In those works, the non-relativistic quark model expression of Eq.~(\ref{eq:expaxffcahn2}) in terms of a single FF $A$ has been assumed, 
and the resulting FF has been parameterized by a dipole~:
\begin{eqnarray}
\frac{A(Q_1^2, 0)}{A(0, 0) } 
= \frac{1}{\left( 1 + Q_1^2/ \Lambda_A^2 \right)^2}\, ,
\label{eq:axffcahnexp}
\end{eqnarray}
where $\Lambda_A$ is a dipole mass. By fitting the resulting expression of Eq.~(\ref{eq:expaxffcahn2}) to experiment (for which $\varepsilon_1 \approx 1$, and for a $Q_1^2$ range which extends up to 6~GeV$^2$), one can then extract the parameters $\tilde \Gamma_{\gamma \gamma}$ and $\Lambda_1$. 
Table~\ref{tab_ax} shows the present experimental status of the equivalent 
$2 \gamma$ decay widths of the axial-vector mesons $f_1(1285)$, and $f_1(1420)$, which we use in this work. 

\begin{table}[h]
{\centering \begin{tabular}{|c|c|c|c|c|c|}
\hline
& $m_A$   & $\tilde \Gamma_{\gamma \gamma} $  & $\Lambda_A$   \\
&  [MeV] &  [keV]  & [MeV] \\
\hline 
\quad $f_1 (1285)$  \quad & \quad $1281.8 \pm 0.6$ \quad   & \quad  $3.5 \pm 0.8 $  \quad  & \quad  $1040 \pm 78 $  \quad   \\
\quad $f_1 (1420)$ \quad  & \quad $1426.4 \pm 0.9$  \quad  & \quad  $ 3.2 \pm 0.9 $ \quad  & \quad  $ 926 \pm 78 $ \quad \\
\hline
\end{tabular}\par}
\caption{Present values~\cite{PDG} of the 
$f_1(1285)$  meson~ and $f_1(1420)$ meson masses $m_A$, their 
equivalent $2 \gamma$ decay widths $\tilde \Gamma_{\gamma \gamma}$, defined according to Eq.~(\ref{a2gwidth}), as well as their 
dipole masses $\Lambda_A$ entering the FF of Eq.~(\ref{eq:axffcahnexp}). 
For $\tilde \Gamma_{\gamma \gamma}$, we use the experimental results from the L3 Collaboration~:  
$f_1(1285)$ from Ref.~\cite{Achard:2001uu}, 
$f_1(1420)$ from Ref.~\cite{Achard:2007hm}. 
Note that for the $f_1(1420)$ state, only the branching ratio $\tilde \Gamma_{\gamma \gamma} \times \Gamma_{K \bar K \pi} / \Gamma_{total}$ is measured so far, which we use as a lower limit on $\tilde \Gamma_{\gamma \gamma}$. }
\label{tab_ax}
\end{table}

\subsection{Tensor mesons}

The process 
$\gamma^\ast (q_1, \lambda_1) + \gamma^\ast(q_2, \lambda_2) \to {\cal T}(\Lambda)$, 
describing the transition from an initial state of two  
virtual photons to a tensor meson ${\cal T}$ ($J^{PC} = 2^{++}$) with mass $m_T$ and 
helicity $\Lambda = \pm 2, \pm 1, 0$ (defined along the direction of $\vec q_1$), is described by 
five independent structures~\cite{Poppe:1986dq, Schuler:1997yw}, and can be parameterized as~:
\begin{eqnarray}
{\cal M}(\lambda_1, \lambda_2; \Lambda) &=& e^2 \, 
\varepsilon_\mu(q_1, \lambda_1) \, \varepsilon_\nu(q_2, \lambda_2) \, 
\varepsilon^\ast_{\alpha \beta}(p_f, \Lambda) \, \nonumber \\
&\times& \left\{ 
 \left[ R^{\mu \alpha} (q_1, q_2) R^{\nu \beta} (q_1, q_2) 
+ \frac{s}{8 X} \, R^{\mu \nu}(q_1, q_2) 
(q_1 - q_2)^\alpha \, (q_1 - q_2)^\beta \right] \, \frac{\nu}{m_T} \,T^{(2)}(Q_1^2, Q_2^2)
\right. \nonumber \\
&&+ \, R^{\nu \alpha}(q_1, q_2) (q_1 - q_2)^\beta  \left( q_1^\mu + \frac{Q_1^2}{\nu} q_2^{\mu} \right) 
\, \frac{1}{m_T} \, T^{(1)}(Q_1^2, Q_2^2) \nonumber \\
&&+ R^{\mu \alpha}(q_1, q_2) (q_2 - q_1)^\beta   \left( q_2^\nu + \frac{Q_2^2}{\nu} q_1^{\nu} \right) 
\,  \frac{1}{m_T} \, T^{(1)}(Q_2^2, Q_1^2)
\nonumber \\ 
&&+ \, R^{\mu \nu}(q_1, q_2) (q_1 - q_2)^\alpha \, (q_1 - q_2)^\beta \, \frac{1}{m_T} \, 
 T^{(0, T)}(Q_1^2, Q_2^2) \, \nonumber \\
&&\left. + \,  \left( q_1^\mu + \frac{Q_1^2}{\nu} q_2^{\mu} \right) \left( q_2^\nu + \frac{Q_2^2}{\nu} q_1^{\nu} \right)  
(q_1 - q_2)^\alpha  (q_1 - q_2)^\beta 
\, \frac{1}{m_T^3} \, T^{(0, L)}(Q_1^2, Q_2^2)
\right\},
\label{eq:tensorff}
\end{eqnarray}
where $\varepsilon_{\alpha \beta}(p_f, \Lambda)$ is the polarization tensor for the tensor meson with 
four-momentum $p_f$ and helicity $\Lambda$.
Furthermore in Eq.~(\ref{eq:tensorff}) $T^{(\Lambda)}$ are the $\gamma^\ast \gamma^\ast \to {\cal T}$ 
transition form factors, for tensor meson helicity $\Lambda$. 
 For the case of helicity zero, there are two form factors depending on whether both photons are transverse (superscript $T$) or longitudinal (superscript $L$). 
 
From Eq.~(\ref{eq:tensorff}), we can easily calculate the different helicity amplitudes as~:
\begin{eqnarray}
&&{\cal M}(\lambda_1 = +1, \lambda_2 = -1; \Lambda = +2) =  
{\cal M}(\lambda_1 = -1, \lambda_2 = +1; \Lambda = -2) = e^2 \, \frac{\nu}{m_T} \, T^{(2)}(Q_1^2, Q_2^2) \, , \nonumber \\
&&{\cal M}(\lambda_1 = 0, \lambda_2 = +1; \Lambda = -1) = - e^2 \, Q_1 \,  \frac{1}{\sqrt{2}} \, \left( \frac{2 X}{\nu m_T^2} \right)
\, T^{(1)}(Q_1^2, Q_2^2) \, , \nonumber \\
&&{\cal M}(\lambda_1 = -1, \lambda_2 = 0; \Lambda = -1) =  e^2 \, Q_2 \, \frac{1}{\sqrt{2}} \, \left( \frac{2 X}{\nu m_T^2} \right) 
\, T^{(1)}(Q_2^2, Q_1^2) \, , \nonumber \\
&&{\cal M}(\lambda_1 = +1, \lambda_2 = +1; \Lambda = 0) =  
{\cal M}(\lambda_1 = -1, \lambda_2 = -1; \Lambda = 0) = - e^2 \, \sqrt{\frac{2}{3}} \, \left( \frac{4 X}{m_T^3} \right) \, 
T^{(0, T)}(Q_1^2, Q_2^2) \, , \nonumber \\
&&{\cal M}(\lambda_1 = 0, \lambda_2 = 0; \Lambda = 0) =  - e^2 \, Q_1 Q_2 \, \sqrt{\frac{2}{3}} \, 
 \left( \frac{4 X^2}{\nu^2 m_T^5} \right) \, T^{(0, L)}(Q_1^2, Q_2^2)\, . 
\end{eqnarray}
The transverse FFs $T^{(2)}$ and $T^{(0, T)}$ at $Q_1^2 = Q_2^2 = 0$ describe the  
two-photon decay widths of the tensor meson with helicities $\Lambda = 2$ and ~$\Lambda = 0$  respectively~:

\begin{eqnarray}
\Gamma_{\gamma \gamma}  \left({\cal T}( \Lambda = 2) \right) &=& \frac{\pi \alpha^2}{4}  \, m_T \, \frac{1}{5} \, 
| T^{(2)} (0,0) |^2 \, , \nonumber \\
\Gamma_{\gamma \gamma} \left({\cal T}(\Lambda = 0) \right) &=& \frac{\pi \alpha^2}{4} \, m_T   \, \frac{2}{15} \,  
| T^{(0, T)} (0,0) |^2 \, .
\label{t2gwidth}
\end{eqnarray}

In this work, we study the sum rules involving cross sections for one real photon and one virtual photon.  
For one quasi-real photon ($Q_2^2 \to 0$), we can obtain from the above helicity amplitudes and using Eq.~(\ref{eq:abs}) the tensor meson contributions to the response functions of Eq.~(\ref{eq:vcross}) as~:
\begin{eqnarray}
\left[ \sigma_2 \right]_{Q_2^2 = 0}  
&=& \delta(s - m_T^2) \, 16 \, \pi^2 \frac{5 \, \Gamma_{\gamma \gamma}({\cal T}(\Lambda = 2))}{m_T} \, 
\left( 1 + \frac{Q_1^2}{m_T^2} \right) 
\, \left[ \frac{T^{(2)}(Q_1^2, 0)}{T^{(2)}(0, 0)} \right]^2 \, ,
\nonumber \\
\left[ \sigma_0 \right]_{Q_2^2 = 0}  
&=& \delta(s - m_T^2) \, 16 \, \pi^2 \, \frac{5 \, \Gamma_{\gamma \gamma}({\cal T}(\Lambda = 0))}{m_T} \, 
\left( 1 + \frac{Q_1^2}{m_T^2} \right)^3 
\, \left[ \frac{T^{(0, T)}(Q_1^2, 0)}{T^{(0, T)}(0, 0)} \right]^2 \, ,
\nonumber \\
\left[ \sigma_\parallel \right]_{Q_2^2 = 0}  &=& \left[ \frac{1}{2} \sigma_2 + \sigma_0 \right]_{Q_2^2 = 0} \, ,
\nonumber \\
\left[ \sigma_\perp \right]_{Q_2^2 = 0}  &=& \left[ \frac{1}{2} \sigma_2 \right]_{Q_2^2 = 0}  \, ,
\nonumber \\
\left[ \sigma_{LT} \right]_{Q_2^2 = 0} &=& \delta(s - m_T^2) \, 8 \, \pi^3 \alpha^2 \, \frac{Q_1^2}{m_T^2} \,  \left( 1 + \frac{Q_1^2}{m_T^2} \right) \,
 \left[ T^{(1)}(Q_1^2,0) \right]^2\, ,
\nonumber \\
\left[ \frac{1}{Q_1 Q_2} \tau_{TL} \right]_{Q_2^2 = 0} &=& \delta(s - m_T^2) \,8 \, \pi^3 \alpha^2 \, 
\frac{1}{m_T^2} \,  \left( 1 + \frac{Q_1^2}{m_T^2} \right)  \nonumber \\ 
&\times& \left\{ \frac{2}{3} \left( 1 + \frac{Q_1^2}{m_T^2} \right)^2  T^{(0, T)}(Q_1^2, 0)  \, T^{(0, L)}(Q_1^2, 0) - \frac{1}{2} T^{(1)}(Q_1^2,0) \, T^{(1)}(0, Q_1^2) \right\} \, ,
\nonumber \\
\left[ \frac{1}{Q_1 Q_2} \tau^a_{TL} \right]_{Q_2^2 = 0} &=& \delta(s - m_T^2) \,8 \, \pi^3 \alpha^2 \, 
\frac{1}{m_T^2} \,  \left( 1 + \frac{Q_1^2}{m_T^2} \right)  \nonumber \\ 
&\times& \left\{ \frac{2}{3} \left( 1 + \frac{Q_1^2}{m_T^2} \right)^2  T^{(0, T)}(Q_1^2, 0)  \, T^{(0, L)}(Q_1^2, 0) + \frac{1}{2} T^{(1)}(Q_1^2,0) \, T^{(1)}(0, Q_1^2) \right\} \, ,
\nonumber \\
\left[ \sigma_{LL} \right]_{Q_2^2 = 0} &=& 0\, . 
\label{eq:tcross}
\end{eqnarray}

\end{document}